# Algorithmic collusion: A critical review

**Florian E. Dorner**

Institute of Science, Technology and Policy, ETH Zurich

florian.dorner@istp.ethz.ch

**Abstract**

The prospect of collusive agreements being stabilized via the use of pricing algorithms is widely discussed by antitrust experts and economists. However, the literature is often lacking the perspective of computer scientists, and seems to regularly overestimate the applicability of recent progress in machine learning to the complex coordination problem firms face in forming cartels. Similarly, modelling results supporting the possibility of collusion by learning algorithms often use simple market simulations which allows them to use simple algorithms that do not produce many of the problems machine learning practitioners have to deal with in real-world problems, which could prove to be particularly detrimental to learning collusive agreements. After critically reviewing the literature on algorithmic collusion, and connecting it to results from computer science, we find that while it is likely too early to adapt antitrust law to be able to deal with self-learning algorithms colluding in real markets, other forms of algorithmic collusion, such as hub-and-spoke arrangements facilitated by centralized pricing algorithms might already warrant legislative action.

# Introduction

Firms' pricing decisions are increasingly automated by pricing algorithms that use market data to adjust prices[1]. Pricing algorithms allow firms to act more efficiently by monitoring large amounts of data and quickly updating prices for a wide range of products. Pricing algorithms might be particularly popular in the internet economy because of the wide availability of information about both customers' buying behaviour and competitors' pricing habits. Indeed, researchers have identified hundreds of sellers in the Amazon Marketplace showing strong indications of algorithmic pricing[2], and a tenfold increase in the rate of price changes by Amazon itself between December 2012 and 2013[3] is strongly indicative of an adoption of pricing algorithms by Amazon in that time frame.

Pricing algorithms could increase consumer welfare via cost reductions in sales departments[4], quicker reactions of firms to improved supply conditions[5], and lower search costs for consumers[6] as their willingness to pay is estimated more accurately. However, there are at least two principal mechanisms by which the widespread adoption of pricing algorithms could harm consumer welfare: The first, and perhaps less controversial is personalized pricing, or put less euphemistically, price discrimination[7]: Better estimates of individual consumers' willingness to pay, facilitated by larger amounts of consumer data and better predictive algorithms, can allow firms to adopt their prices accordingly[8], and to capture large amounts of consumer welfare. The second mechanism and main focus of this paper is algorithmic collusion: As Ezrachi and Stucke[9] argue, the use of pricing algorithms could: 1) make it easier for firms to implement collusive agreements and set supracompetitive prices in markets where such agreements are already possible, for example by automating the monitoring of adherence to such agreements, 2) enable the formation of such agreements in new markets where collusion was previously infeasible, for example by using the fast reaction time of algorithms to instantly punish firms that undercut prices such that undercutting ceases to be a viable strategy, or 3) facilitate the "tacit" formation of implicit collusive agreements, that does not require any explicit communication that could be used as evidence against the participants in antitrust cases. For example, this could happen as self-learning algorithms deployed by different firms to maximize individual profits learn to cooperate with each other to achieve their goals. Unlike with price discrimination, algorithmic collusion does not "just" affect redistribution[10], but also negatively affects overall efficiency, as colluding firms will supply less than they would supply in a competitive market, to keep prices at a higher level.

---

[1] Assad, Stephanie, et al. "Algorithmic pricing and competition: Empirical evidence from the German retail gasoline market." (2020).

[2] Chen, Le, Alan Mislove, and Christo Wilson. "An empirical analysis of algorithmic pricing on amazon marketplace." *Proceedings of the 25th international conference on World Wide Web*. 2016.

[3] Competition and Markets Authority. "Pricing algorithms; Economic working paper on the use of algorithms to facilitate collusion and personalised pricing" (2018) https://assets.publishing.service.gov.uk/government/uploads/system/uploads/attachment_data/file/746353/Algorithms_econ_report.pdf

[4] Mehra, Salil K. "Antitrust and the robo-seller: Competition in the time of algorithms." *Minn. L. Rev.* 100 (2015): 1323.

[5] OECD, "Algorithms and Collusion: Competition Policy in the Digital Age" (2017) www.oecd.org/competition/algorithms-collusion-competition-policy-in-the-digital-age.htm

[6] Siciliani, Paolo. "Tackling Algorithmic-Facilitated Tacit Collusion in a Proportionate Way." Journal of European Competition Law & Practice 10.1 (2019): 31-35.

The third type of algorithmic collusion involving self-learning approaches from artificial intelligence (AI) might be the most problematic, as both the legality of and liability around collusive agreements formed by AI algorithms without explicit instruction to collude are controversial to say the least. Furthermore, AI approaches are also likely to play an important role for the second type of algorithmic collusion, which is in fact a superset of the third: The more complex a market gets, the harder it becomes for programmers to clearly define and "hard-code" collusive agreements, such that AI might be necessary for enabling algorithmic collusion in many markets. For example, AI might be needed to obtain a better estimates of the demand other firms' face, in order to avoid costly punishment of others' reactions to demand shifts that maintain collusive profits, as well as the resulting potential for price wars.

While there indeed seem to be firms advertising the anticompetitive capabilities of their pricing algorithms[11], the plausibility of AI-enabled collusion has been called into question by several scholars: While Nicolas Petit still hedges his comparison of the literature on antitrust and AI to science fiction by stating that science fiction unearths fascinating research hypotheses[12], Ulrich Schwalbe, after reviewing the relevant literature on game theory and reinforcement learning (RL), a form of AI used to train agents, concludes that "algorithmic collusion currently seems far more difficult to achieve than legal scholars have often assumed and is thus not a particularly relevant competitive concern at present."[13]. Similarly, a 2019 joint report by the german and french antitrust authorities preliminarily calls into question whether the theoretical and experimental modelling results on algorithmic collusion by self-learning agents provide strong evidence for the collusive potential of such algorithms in real markets[14]. The likely current inadequacy of AI algorithms at achieving collusive outcomes in complex markets is also highlighted by a recent research program on cooperative AI involving researchers from DeepMind[15][16], the subsidiary of Google's parent company Alphabet, that used AI to build the first computer to beat the human world champion in the game of Go[17]. The researchers identify many roadblocks for cooperation between AI systems in mixed-motive games that are neither entirely zero-sum, nor fully cooperative: These obstacles include: 1) Algorithms' ability to understand other algorithms, which seems to often be assumed as trivial by researches in the field of algorithmic collusion such as Michal Gal[18], but can be challenging, especially because of issues with recursion[19], 2) algorithms' ability to communicate with each other, which might require some established common grounding of communication, 3) algorithms ability to commit to cooperation, even if there might be naively rational reasons to defect instead, and 4) the lack of institutions and norms bounding behaviour, both of which arguably play an important role for human cooperation. As collusion can easily be framed as a form of (antisocial) cooperation between firms, and as firms in a market clearly have mixed motives[20], some or all of these problems are likely to apply to collusion as much as they do to other forms of cooperation.

---

[7] Mankiw, N., and Mark Taylor. *Economics*. 5th ed., Cengage Learning EMEA (2020).
[8] Bergemann, Dirk, Alessandro Bonatti, and Tan Gan. "The Economics of Social Data." *arXiv preprint arXiv:2004.03107* (2020).
[9] Ezrachi, Ariel, and Maurice E. Stucke. "Artificial intelligence & collusion: When computers inhibit competition." *U. Ill. L. Rev.* (2017): 1775.
[10] Strictly speaking, price discrimination can even increase economic efficiency, as argued in Mankiw and Taylor (supra): Deadweight loss can be reduced with perfect price discrimination. as some customers who would not have bought the product at the profit-maximizing price still get to buy it when prices are adapted to their willingness to pay as long as that is about the firm's marginal cost.
[11] Ezrachi, Ariel, and Maurice E. Stucke. "Sustainable and unchallenged algorithmic tacit collusion." *Nw. J. Tech. & Intell. Prop.* 17 (2019): 217.

While demonstrating that collusion by AI systems is not very likely to be feasible at the current time, the demonstrated interest of high-profile machine learning researchers in cooperative AI capabilities clearly demonstrates that this might change in the future. This is especially the case because it seems likely that there would not be any fundamental barriers to algorithms learning to cooperate, given that humans can do so in a wide range of situations. Also, as the researchers point out, there are likely strong reasons to develop AI

---

[12] Petit, Nicolas. "Antitrust and artificial intelligence: a research agenda." *Journal of European Competition Law & Practice* 8.6 (2017): 361-362.
[13] Schwalbe, Ulrich. "Algorithms, machine learning, and collusion." *Journal of Competition Law & Economics* 14.4 (2018): 568-607.
[14] Bundeskartellamt and Autorité de la concurrence, "Algorithms and Competition" (2019) https://www.bundeskartellamt.de/SharedDocs/Publikation/EN/Berichte/Algorithms_and_Competition_Working-Paper.pdf;jsessionid=1CDD8FD2F88186AAF0BA5C72206E0787.2_cid387?__blob=publicationFile&v=5
[15] Dafoe, Allan, et al. "Open problems in cooperative AI." *arXiv preprint arXiv:2012.08630* (2020).
[16] Dafoe, Allan, et al. "Cooperative AI: machines must learn to find common ground." *Nature* 593.7857 (2021): 33-36.
[17] Silver, David, et al. "Mastering the game of go without human knowledge." *nature* 550.7676 (2017): 354-359.
[18] Gal, Michal S. "Algorithms as illegal agreements." *Berkeley Tech. LJ* 34 (2019): 67.
[19] An algorithms beliefs about another algorithm's beliefs might need to include the later algorithm's beliefs about the first algorithm's beliefs, which can make it difficult for both algorithms to form correct beliefs about each other, as this requires consistency between their own beliefs and their own beliefs about what the other algorithm beliefs their own beliefs are.
[20] If firms were playing zero-sum games, collusion would not be a problem and we would see perfect competition at best, or widespread sabotage at worse. If they were playing fully cooperative games, collusion would consistently occur, with or without algorithms.

systems that are able to cooperate with other AI systems and humans instead of engaging in costly conflicts, both for prosocially motivated researchers and profit oriented companies. Even if the prosocial motivation was not a given, research on cooperative AI would likely continue because of economic incentives alone, such that continued attention to the problem of AI-based algorithmic collusion is likely warranted despite its unlikelihood in the very near term. In particular, a better understanding of the obstacles to AI-based collusion could help to identify "canaries in the coal mine"[21], indicators of an imminent threat to competition that warrants quick legislative action. This would be particularly useful, as a more cautious approach would likely be preferable if no such indicators were observed, as firms might also have strong competitive incentives to adopt pricing algorithms, for example to cut costs, and as it can better take into account specific details of the most relevant subset of AI approaches used, once AI based collusion becomes possible. In words attributed the European Commissioner for Competition, Margrethe Vestager:

*"It is true that the idea of automated systems getting together and reaching a meeting of minds is still science fiction. [...] But we do need to keep a close eye on how algorithms are developing. [...] So that when science fiction becomes reality, we are ready to deal with it."*[22].

In addition, AI based collusion is not fundamentally different from other forms of algorithmic collusion, but rather a point on a spectrum. As such, there are ample synergies between a better understanding of AI-based collusion and simpler forms of algorithmic collusion, the latter of which might already have immediate relevance, even if it probably poses less severe problems to antitrust law.

In the following, we begin our investigation of algorithmic collusion by broadly defining collusion and giving an overview of the relevant antitrust regulations from both the US and the EU. Then, we review important determinants for collusive outcomes as laid out in the economics literature, and critically analyze, how pricing algorithms could simplify collusion or prevent the detection of collusive agreements. After reviewing the existing arguments and evidence for and against the severity of algorithmic collusion, we conclude by discussing relevant gaps regarding the prevention of algorithmic collusion, as well as potential solutions.

## Defining collusion

To analyze algorithmic collusion it is first necessary to understand collusion, which requires a working definition: Collusion can be defined as a coordinated strategy firms in a market employ to increase their joint profits, usually at the expense of consumers via setting supracompetitive prices. This usually requires firms to first agree on such a strategy and then stabilize it via monitoring others' adherence to the strategy and punishing deviations[23]. In explicit collusion, this involves explicit communication about the joint strategy, while tacit (or implicit) collusion involves firms adapting their behaviour towards a collusive strategy without the need for explicit communication[24]. For example, tacit collusion might be reached if two firms in a duopoly independently decide to "retaliate" against price cuts by the other via temporarily lowering their own price by an even large amount. Once both firms are aware that they are both using this strategy, neither of them might have an incentive to lower prices below the monopoly level, because the other firms' undercutting of their price would otherwise lower their profits.

---

[21] Cremer, Carla Zoe, and Jess Whittlestone. "Artificial Canaries: Early Warning Signs for Anticipatory and Democratic Governance of AI." *International Journal of Interactive Multimedia & Artificial Intelligence* 6.5 (2021).
[22] Quote extracted from Klein, Timo. "Autonomous algorithmic collusion: Q-learning under sequential pricing." *Amsterdam Law School Research Paper* 2018-15 (2019): 2018-05.
[23] OECD, "Algorithms and Collusion: Competition Policy in the Digital Age" (2017) www.oecd.org/competition/algorithms-collusion-competition-policy-in-the-digital-age.htm

This example also illustrates the need for punishing deviations for the maintenance of collusive strategies. In fact, some definitions of collusion require the presence and mutual awareness of punishment for firms deviating from the supracompetitive price[25]. The focus on punishment highlights an interesting distinction: Is punishment in form of undercutting others' prices in response to a deviation required to incentivize others to collude costly in the short term or not? Costly punishment only makes sense a part of a deliberate long term strategy to bring other firms to participate in a collusive strategy. If anticipated, costly punishment is likely to be highly effective as it also raises the costs others incur for their deviations. For example, the "grim trigger" strategy[26] punishes deviations with a permanent price cut down to the marginal cost, eliminating most or all of the profits for the involved firms. However, if deviations still occur, perhaps as the severity of punishment was underestimated, more costly punishment is bad for both firms. On the other hand, if undercutting in response to deviations is not costly relative to the short-term profit-maximizing price, given the other firms' prices, it is not even clear whether it can be considered punishment, as it is just the selfish, competitive response. Notably, such non-costly punishment still creates an incentive for others not to reduce their prices, but this incentive can be a lot weaker than with costly punishment, such that the incentive to undercut prices might be stronger if firms are not very farsighted[27]. In particular, most fully competitive markets do exhibit the described non-costly "punishment" dynamic[28]. As firms are usually unable to affect the market characteristics determining the extent of non-costly punishment, only the presence of costly punishment can constitute relevantly strong evidence for firms' attempts to tacitly collude. However, as non-costly punishment might sometimes be sufficient for facilitating tacit collusion, other markers of tacit collusion such as restraint from lowering prices despite obvious short term gains to such a price drop might need to be considered absent costly punishment.

## US antitrust law

As antitrust law usually does not prohibit collusion per se, no effort is made to legally define collusion. Instead, the law usually focuses on the means used by firms to achieve collusive outcomes. Interestingly this focus is not on punishment, but on "unreasonable" agreements to employ a collusive strategy. Indeed, antitrust law often requires proof that firms did not act

---

[24] Monopolkommission, "Wettbewerb 2018 XXII. Hauptgutachten der Monopolkommission gemäß §44 Abs.1 Satz 1 GWB" (2018) https://monopolkommission.de/images/HG22/HGXXII_Gesamt.pdf
[25] Harrington, Joseph E. "Developing competition law for collusion by autonomous artificial agents." *Journal of Competition Law & Economics* 14.3 (2018): 331-363.
[26] Miklós-Thal, Jeanine, and Catherine Tucker. "Collusion by algorithm: Does better demand prediction facilitate coordination between sellers?." *Management Science* 65.4 (2019): 1552-1561.
[27] Formally, if they do not discount future earnings very slowly.
[28] How would a competitive equilibrium be reached if not by firms repeatedly undercutting each other?

independently and had a so-called "meeting of the minds"[29]. In the US, section 1 of the Sherman Act[30] states:

*"Every contract, combination in the form of trust or other- wise, or conspiracy, in restraint of trade or commerce among the several States, or with foreign nations, is hereby declared to be illegal. Every person who shall make any such contract or engage in any such combination or conspiracy, shall be deemed guilty of a misdemeanor, and, on conviction thereof, shall be punished by fine not exceeding five thousand dollars, or by imprisonment not exceeding one year, or by both said punishments, at the discretion of the court. [...]"*

According to Joseph Harrington[31], the formulation of *"contract, combination in the form of trust or other- wise, or conspiracy"* is mostly interpreted as synonymous with "agreement", while "*restraint of trade or commerce*" is interpreted as "limiting competition". Then again "agreement" has been interpreted as "unity of purpose or a common design and understanding, or a meeting of minds in an unlawful arrangement"[32] or "a conscious commitment to a common scheme designed to achieve an unlawful objective."[33] in important antitrust decisions. This focus on agreements rather than collusion itself means that courts are likely to require evidence of such an agreement, usually in the form of evidence for explicit communication[34]; As stated by former circuit judge Richard Posner: *"But section 1 of the Sherman Act, under which the suit had been brought, does not require sellers to compete; it just forbids their agreeing or conspiring not to compete. So as the Court pointed out, a complaint that merely alleges parallel behavior alleges facts that are equally consistent with an inference that the defendants are conspiring and an inference that the conditions of their market have enabled them to avoid competing without having to agree not to compete."*. This focus on evidence for an agreement rather than for collusion is of practical nature: Associate Justice of the Supreme Court Stephen Breyer states: *"Courts have noted that the Sherman Act prohibits agreements, and they have almost uniformly held, at least in the pricing area, that such individual pricing decisions (even when each firm rests its own decision upon its belief that competitors will do the same) do not constitute an unlawful agreement under section 1 of the Sherman Act. [...] That is not because such pricing is desirable (it is not), but because it is close to impossible to devise a judicially enforceable remedy for "interdependent" pricing. How does one order a firm to set its prices without regard to the likely reactions of its competitors?"*[35]. However, the other key piece of US antitrust legislation[36] the Federal Trade Commission (FTC) act[37] is scoped more broadly

---

[29] OECD, "Algorithms and Collusion: Competition Policy in the Digital Age" (2017) www.oecd.org/competition/algorithms-collusion-competition-policy-in-the-digital-age.htm

[30] Sherman Act (Sec. 1). Quote extracted from https://www.ourdocuments.gov/doc.php?flash=false&doc=51&page=transcript accessed on the 12th of July 2021.

[31] Harrington, Joseph E. "Developing competition law for collusion by autonomous artificial agents." *Journal of Competition Law & Economics* 14.3 (2018): 331-363.

[32] Am. Tobacco Co. v. United States, 328 U.S. 781 S. Ct. 1125 (1946).

[33] Monsanto Co. v. Spray-Rite Serv. Corp., 465 U.S. 752 S. Ct. 1464 (1984).

[34] Page, William H. "Communication and Concerted Action." *Loy. U. Chi. LJ* 38 (2006): 405.

[35] Clamp-All Corp. v. Cast Iron Soil Pipe Inst., 851 F.2d 478 (1st Cir. 1988).

[36] https://www.ftc.gov/tips-advice/competition-guidance/guide-antitrust-laws/antitrust-laws Accessed on the 12th of July 2021.

[37] FTC act (Sec. 5 (1,2)). Quote extracted from https://www.ftc.gov/sites/default/files/documents/statutes/federal-trade-commission-act/ftc_act_incorporatingus_safe_web_act.pdf accessed on the 12th of July 2021.

and does not rely on the notion of agreement. Instead of focussing on specific rules, it relies on the more flexible principle of "fairness"[38]:

*"Unfair methods of competition in or affecting commerce, and unfair or deceptive acts or practices in or affecting commerce, are hereby declared unlawful"* and *"The Commission is hereby empowered and directed to prevent persons, partnerships, or corporations, except [...] from using unfair methods of competition in or affecting commerce and unfair or deceptive acts or practices in or affecting commerce. [...]"*

Former Circuit Judge Walter R. Mansfield interpreted this as follows: *"In our view, before business conduct in an oligopolistic industry may be labelled "unfair" within the meaning of Sec. 5 a minimum standard demands that, absent a tacit agreement, at least some indicia of oppressiveness must exist such as (1) evidence of anticompetitive intent or purpose on the part of the producer charged, or (2) the absence of an independent legitimate business reason for its conduct."*[39]. In other words, evidence for agreement might not always be required to declare collusion illegal, but the alternatives of proving intent or a lack of legitimate business reasons for a firms' strategies might be similarly hard to prove.

## EU antitrust law

Unlike the Sherman Act, the more recently implemented EU law directly talks about agreements: Article 101 of the Treaty on the Functioning of the European Union (TFEU)[40] formerly known as Article 81 of the Treaty establishing the European Community or Article 85 of the Treaty establishing the European Economic Community [41] states:

*"The following shall be prohibited as incompatible with the internal market: all agreements between undertakings, decisions by associations of undertakings and concerted practices which may affect trade between Member States and which have as their object or effect the prevention, restriction or distortion of competition within the internal market, and in particular those which: (a) directly or indirectly fix purchase or selling prices or any other trading conditions [...]"*. However, exceptions exist: *"[...] The provisions of paragraph 1 may, however, be declared inapplicable in the case of: any agreement or category of agreements between undertakings [...] which contributes to improving the production or distribution of goods or to promoting technical or economic progress, while allowing consumers a fair share of the resulting benefit, and which does not: a) impose on the undertakings concerned restrictions which are not indispensable to the attainment of these objectives; b) afford such undertakings the possibility of eliminating competition in respect of a substantial part of the products in question."*

---

[38] OECD, "Algorithms and Collusion: Competition Policy in the Digital Age" (2017) www.oecd.org/competition/algorithms-collusion-competition-policy-in-the-digital-age.htm
[39] Ethyl Corp. v. Fed. Trade Comm'n, 729 F.2d 128 (2d Cir. 1984).
[40] TFEU, Article 101
https://eur-lex.europa.eu/LexUriServ/LexUriServ.do?uri=CELEX:12008E101:EN:HTML accessed on July 12th 2021
[41] EC Treaty, Article 81
https://eur-lex.europa.eu/LexUriServ/LexUriServ.do?uri=CELEX:12002E081:EN:HTML accessed on July 13th 2021

Again, evidence for explicit communication plays an important role in related court rulings: *"[...] where the Commission's reasoning is based on the supposition that the facts established in its decision cannot be explained other than by concertation between the undertakings, it is sufficient for the applicants to prove circumstances which cast the facts established by the Commission in a different light and thus allow another explanation of the facts to be substituted for the one adopted by the Commission. However, the Court specified that that case-law was not applicable where the proof of concertation between the undertakings is based not on a mere finding of parallel market conduct but on documents which show that the practices were the result of concertation [...]"* [42] However, explicit communication might not be necessary to prove illegal agreements under european law, if the involved firms are unable to provide an alternative explanation for their seemingly collusive behaviour: *"Although parallel behaviour may not by itself be identified with a concerted practice, it may however amount to strong evidence of such a practice if it leads to conditions of competition which do not correspond to the normal conditions of the market, having regard to the nature of the products, the size and number of the undertakings, and the volume of the said market. This is especially the case if the parallel conduct is such as to enable those concerned to attempt to stabilize prices at a level different from that to which competition would have led"*[43]. In addition, like in the US, EU antitrust law also features more broadly scoped legislation against the abuse of market power, in form of TFEU, Article 102[44]:

*"Any abuse by one or more undertakings of a dominant position within the internal market or in a substantial part of it shall be prohibited as incompatible with the internal market in so far as it may affect trade between Member States. Such abuse may, in particular, consist in: (a) directly or indirectly imposing unfair purchase or selling prices or other unfair trading conditions [...]"*

As for the FTC act, the use reference to "fairness" replaces the focus on agreement, making TFEU 102 both more vague and more flexible than TFEU 101 or the Sherman Act.

## Is algorithmic collusion plausible?

Now that we have defined collusion and delineated how antitrust law attempts to prohibit it, we narrow our focus down to algorithmic collusion. We begin by presenting three stylized scenarios for how algorithms could enable collusion. Then, we provide an overview of risk factors that determine the likelihood of collusion, and review, which characteristics of algorithms might make algorithmic collusion more easy to achieve or hard to detect, compared to human collusion. Lastly, we briefly review the empirical evidence for both the prevalence of algorithmic pricing and algorithmic collusion, as well as results on algorithmic collusion from oligopoly models.

---

[42]Case T-442/08, CISAC v Comm'n (2013)
https://curia.europa.eu/juris/document/document.jsf?text=&docid=136261&pageIndex=0&doclang=en&mode=req&dir=&occ=first&part=1&cid=357698 accessed on July 13th 2021
[43]Case 48/69, Imperial Chemical Industries Ltd. v. Comm'n, (1972)
 https://eur-lex.europa.eu/legal-content/en/ALL/?uri=CELEX:61969CJ0048
[44]TFEU, Article 101
https://eur-lex.europa.eu/LexUriServ/LexUriServ.do?uri=CELEX:12008E102:EN:HTML accessed on July 12th 2021

## Types of algorithmic collusion

As algorithmic collusion can take different forms, which differ in both their likelihood and the challenges they pose to the the legal system, we differentiate three types of algorithmic collusion to guide the rest of our discussion. While our typology is largely base on Ezrachi and Stucke's seminal paper on algorithmic collusion[45], we combine their "Predictable Agent" and "Digital Eye" scenario into one, as both are based on algorithms colluding without explicit instructions to collude.

### Messenger

In the first scenario, which is similar to the scenario termed "Messenger" by Ezrachi and Stucke, algorithms are used as mere tools used by humans to initiate, monitor and/or automatically enforce classical collusive agreements. For example, firms could have a meeting and jointly decide on setting prices to a fixed supracompetitive level and both use pricing algorithms that take into account the others' price shifts to automatically punish deviations. As algorithms are used as mere tools in this scenario, it might not seem to create any novel legal issues. However, as will become obvious in the rest of this section, many of the arguments for how algorithms could increase the prevalence and decrease the detectability of collusive agreements, in particular tacit ones, already work in this basic scenario.

### Hub-and-Spoke

In the hub-and-spoke scenario, firms in a market form a hub-and-spoke cartel with the third-party developer of a common pricing algorithm. Then, the developer (hub) facilitates collusion between the firms (spokes) by providing them with compatible pricing algorithms that jointly lead to a collusive outcome, or even employing a centralized pricing algorithm that combines sensitive information from participating firms[46]. Note, that sending important information to a third-party service provider might not be very unusual, even absent any intent to collude, given the seeming popularity of cloud-based solutions for business applications. While not involving firms as spokes, companies like Uber can serve as an illustration of the hub-and-spoke problem: By strategically setting the prices for its drivers and preventing drivers from negotiating lower prices themselves, Uber might be able to bring the fees customers pay to its drivers close to the monopoly level, at least in areas without competitors to Uber. In particular, all of the data on supply and demand conditions collected by its drivers could be bundled and used to calculate the profit maximizing price a lot more accurately. Multiple key distinctions might affect who, if anyone, is liable in the hub-and-spoke scenario: Did the spokes know that the hub facilitated collusion, rather than just increasing their profits by other means, such as improved demand predictions[47] [48]? And did the hub use a centralized algorithm or set of algorithms that was explicitly designed to facilitate collusion, or did it merely provide the same (learning) algorithm to all firms, and the algorithm happened to learn collusive strategies when combined with copies of itself?

---

[45] Ezrachi, Ariel, and Maurice E. Stucke. "Artificial intelligence & collusion: When computers inhibit competition." *U. Ill. L. Rev.* (2017): 1775.
[46] Lorenz Marx, Christian Ritz, Jonas Weller, "Liability for outsourced algorithmic collusion: A practical approximation", *Concurrences N° 2-2019, Art. N° 89925, www.concurrences.com* (2019)
[47] Id.
[48] Bundeskartellamt and Autorité de la concurrence, "Algorithms and Competition" (2019)

### Independent algorithms

In the last scenario, pricing algorithms independently deployed by firms eventually come to use collusive strategies without explicitly being programmed to collude. This can either happen through the use of self-learning algorithms, as in Ezrachi and Stucke's "Digital Eye" scenario[49], or as firms iteratively change their algorithms in response to new market conditions generated by their own and others' use of algorithms, without ever explicitly deciding to collude. The second case generalizes Ezrachi and Stucke's "Predictable Agent"[50] scenario, in which the extensive use of pricing algorithms leads to increased market transparency, which can in turn enable more tacit forms of collusion. While the "human in the loop" needed for adapting the algorithm in the second case might simplify some antitrust issues compared to the self-learning algorithm in the first case, the fundamental problem is the same: Algorithms gradually adapt their strategy to market conditions they observe, be it through human adjustment or predefined learning rules. These market conditions include other algorithms' strategies, which are adjusted in turn. This creates a feedback loop that can in some cases move the algorithms' joint strategy towards collusion. Even if the adjustments are performed by a human, they might be explainable as rational responses to the market conditions, rather than attempts at creating collusive outcomes, such that illegality would be hard to establish. In the case of purely self-learning agents, this form of tacit algorithmic collusion would present an even larger problem for antitrust law's ability to prevent collusion, as there might be many strong reasons to deploy self-learning pricing algorithms beyond the potential of collusive outcomes.

## Key determinants of collusion

To evaluate the plausibility and potential impact of our scenarios, we first have a look at various factors the economics literature has identified as relevant for the occurence of collusion.

### Payoffs and risks

The first, and most general factor concerns the gains from successful collusion, the likelihood and size of associated risks, such as other firms reneging on the collusive agreement or fine payments issued by antitrust authorities, as well as gains for reneging oneself. A collusive strategy can only be stable, if firms expect not to gain more from undercutting prices than they lose. This again highlights the importance of punishment: If firms know that their gains from undercutting the collusive price will be more than offset by another firm's (costly) punishment, they are unlikely to deviate[51]. Many factors can affect the payoffs and risks firms face. Amongst others, these include: Demand and cost stability, the size and frequency of purchases[52], buyer power, product homogeneity, and the prevalence of innovation[53]. We refrain from discussing all of these factors in detail, but rather focus on the few that might be most relevant for algorithmic collusion.

---

[49] Ezrachi, Ariel, and Maurice E. Stucke. "Artificial intelligence & collusion: When computers inhibit competition." *U. Ill. L. Rev.* (2017): 1775.
[50] Id.
[51] Harrington, Joseph E. "Developing competition law for collusion by autonomous artificial agents." *Journal of Competition Law & Economics* 14.3 (2018): 331-363.
[52] Deng, Ai. "What Do We Know about Algorithmic Tacit Collusion?." (2018).
[53] Bernhardt, Lea, and Ralf Dewenter. "Collusion by code or algorithmic collusion? When pricing algorithms take over." *European Competition Journal* 16.2-3 (2020): 312-342.

### Amount of competitors and barries to entry

A larger number of companies in a market maces collusion harder to achieve[54] by increasing the complexity and costs[55] for coordination: Finding a common price to agree on can become difficult with many parties, even when explicit communication is involved, and the involvement of multiple parties in such communication is likely to substantially increase the risk of detection. Without communication, tacit punishment strategies might not work very well, as the causal link between a firm's price change and the punishment can be harder to identify, especially if many firms are active in a market. Lastly, depending on the exact market structure, large amounts of firms in a market could also increase the immediate gains from undercutting others[56]. In addition to reducing the amount of market participants, entry barriers might further stabilize collusion by reducing the likelihood that new firms attempt to disrupt a market by undercutting existing firms' collusive prices[57].

### Repeated interactions and speed

As with other forms of cooperation[58], repeated interactions between two firms can foster stable collusion, as punishment from deviations can only be applied after the deviation occurred. The more often firms interact, and the more farsighted they are[59], the larger the threat of future punishment weighs compared to the immediate gains from undercutting prices[60]. While frequent market entries and exits could also have an influence on this factor, the main mechanism of concern is speed: If firms can quickly adjust their prices in response to others' deviations, they increase the effective number of interactions between each other. Viewed from another angle, quick price adjustments reduce the amount of time a firms' price is the lowest after undercutting, and thus the gain from such deviations from collusive behaviour is reduced.

### (Market) Transparency

Supply side market transparency is often cited as a key prerequisite for (tacit) collusion[61] [62] [63]. However, there are at least two different types of market transparency with different mechanisms for how they affect collusion, which are rarely distinguished:

The first concerns the transparency of present market interactions: What are other firms' current prices, and what are the contextual factors that might influence them? Current prices are important, as price deviations can only be punished if they are detected[64]. And knowing contextual factors can prevent firms from accidentally starting price wars in response to price cuts caused by shifts in the market conditions that might be acceptable under a given collusive agreement.

---

[54] Kühn, Kai-Uwe, and Steve Tadelis. "Algorithmic collusion." *Presentation prepared for CRESSE, URL http://www. cresse. info/uploadfiles/2017_sps5_pr2. pdf* (2017).

[55] Bundeskartellamt and Autorité de la concurrence, "Algorithms and Competition" (2019)

[56] If the market exhibits "Winner takes all" dynamics and most customers buy from the cheapest firm (but are distributed evenly between firms if all have the same price), undercutting a collusive price increases the sales from half the market to the whole market with two firms, but from a tenth of the market to the whole market with ten firms.

[57] Bundeskartellamt and Autorité de la concurrence, "Algorithms and Competition" (2019)

[58] Axelrod, Robert, and William Donald Hamilton. "The evolution of cooperation." *science* 211.4489 (1981): 1390-1396.

[59] Meaning that they discount the future at only a small rate.

[60] Bundeskartellamt and Autorité de la concurrence, "Algorithms and Competition" (2019)

[61] OECD, "Algorithms and Collusion: Competition Policy in the Digital Age" (2017) www.oecd.org/competition/algorithms-collusion-competition-policy-in-the-digital-age.htm

[62] Ezrachi, Ariel, and Maurice E. Stucke. "Artificial intelligence & collusion: When computers inhibit competition." *U. Ill. L. Rev.* (2017): 1775.

[63] Interestingly, consumer-facing market transparency might have the opposite effect, as it can increase the gains from undercutting prices, as consumers are better able to identify the best offer.

The second concerns the decision making process firms employ when deciding prices: Price wars might be easier to prevent, if firms know which variables others take into account for making their pricing decision[65]: If one firm's "legitimate" price cut is driven by a variable the other firm is not taking into account, this could be interpreted as a defection and punished, again triggering a price war. More broadly, a strong understanding of another firms' decision process might enable firms to predict their reactions to various price changes[66], which could lead to collusive agreements that neither require explicit communication, nor repeated costly demonstration of punishment.

Communication and Coordination

As the law's focus on explicit communication to establish collusion suggest, communication can play an important role to achieve coordination on a collusive strategy[67] [68]. In particular, (credible) verbal threats to punish deviations can prevent others from cheating on collusive agreements[69]. And while price signals might function as a rudimentary form of communication in some cases[70], it does not seem likely that they would be sufficient for achieving collusion when there is disagreement about how a collusive agreement should look like, for example in market involving asymmetric participants, as discussed later. Naive game theoretic analysis might ignore the importance of communication, as the set of stable collusive strategies or *collusive equilibria* does not necessarily depend on the possibility for communication. However, while two firms playing the grim-trigger strategy is an equilibrium, both firms might easily end up with marginal cost pricing, if only one of them employs it, or even if the firms use a different threshold for what they consider collusive prices. This illustrates that communication can play an important role in *equilibrium selection*[71] [72]. It might also be helpful for reaching *any* equilibrium at all, rather than getting caught in price cycles that yield supracompetitive, but suboptimal payoffs[73]. In addition, communication could be used to facilitate transparency of firms' decision processes and explain "justified" decisions that others might interpret as defection[74].

---

[64] Beneke, Francisco, and Mark-Oliver Mackenrodt. "Remedies for algorithmic tacit collusion." *Journal of Antitrust Enforcement* 9.1 (2021): 152-176.
[65] Gal, Michal S. "Algorithms as illegal agreements." *Berkeley Tech. LJ* 34 (2019): 67.
[66] Ezrachi, Ariel, and Maurice E. Stucke. "Artificial intelligence & collusion: When computers inhibit competition." *U. Ill. L. Rev.* (2017): 1775.
[67] Gal, Michal S. "Algorithms as illegal agreements." *Berkeley Tech. LJ* 34 (2019): 67.
[68] Fonseca, Miguel A., and Hans-Theo Normann. "Explicit vs. tacit collusion—The impact of communication in oligopoly experiments." *European Economic Review* 56.8 (2012): 1759-1772.
[69] Cooper, David J., and Kai-Uwe Kühn. "Communication, renegotiation, and the scope for collusion." *American Economic Journal: Microeconomics* 6.2 (2014): 247-78.
[70] OECD, "Algorithms and Collusion: Competition Policy in the Digital Age" (2017) www.oecd.org/competition/algorithms-collusion-competition-policy-in-the-digital-age.htm
[71] Kaplow, Louis. "On the meaning of horizontal agreements in competition law." *California Law Review* (2011): 683-818. P 797,798.
[72] Bundeskartellamt and Autorité de la concurrence, "Algorithms and Competition" (2019)
[73] For an example of (Q-learning) algorithms engaging in price cycles, see:
Klein, Timo. "Autonomous algorithmic collusion: Q-learning under sequential pricing." *Amsterdam Law School Research Paper* 2018-15 (2019): 2018-05.
[74] Bundeskartellamt and Autorité de la concurrence, "Algorithms and Competition" (2019)

### Credible commitments and ease of enforcement

Threats to punish others for deviating from a collusive strategy can be insufficient for stabilizing that strategy if they are not credible[75]. And credibility might be especially lacking for threats of costly punishment, which would otherwise be most effective: For example, executing a grim trigger strategy maximally hurts the firm executing the punishment as well. Thus, while firms are incentivized to make others belief that they will use maximally costly punishment, they are not actually incentivized to carry out such punishment, except for the purpose of building a reputation as someone who punishes hard. Correspondingly, the ability to credibly commit to a particular strategy, despite local incentives not to use this strategy, can be extremely valuable for facilitating collusion.

One aspect of antitrust law that might be too obvious for most commentators to point out is the prevention of such commitment: As illegal agreements are not enforced by courts, contracts, society's go-to measure to establish commitment to stabilize all kinds of cooperative agreements is out of the question for establishing collusive commitments. Without out access to societies' most important trust building institutions[76], collusion is a lot harder than ordinary, legal cooperation. The credibility of firms' commitments is further undermined by the use of antitrust leniency procedures[77], which incentivize firms to self-report collusive practices they participate in via partial or even full immunity from fines[78], and thus both increase the incentive to leave a collusive agreement, and increase the risk of detection and legal penalties.

### Symmetry

Symmetric competitors with similar product quality and cost structure might have an easier time colluding than firms with more asymmetric market positions. This is because the joint monopoly price, or at least symmetric collusive strategies with equal prices provide an obvious Schelling Point for firms to coordinate around in the symmetric case[79]. If the firms' face asymmetric market conditions, many more potentially stable collusive strategies might exist, but these might not involve the same price for both firms. This way, firms might not only need to decide to collude, but also bargain about how the spoils of collusion are to be distributed, which complicates collusion, especially absent explicit communication.

## Key properties of algorithms

Maureen K. Ohlhausen, the former Acting Chairman of the U.S. Federal Trade Commission: stated: *"Let's just change the terms of the hypothetical slightly to understand why. Everywhere the word "algorithm" appears, please just insert the words "a guy named Bob". Is it ok for a guy named Bob to collect confidential price strategy information from all the participants in a market, and then tell everybody how they should price? If it isn't ok for a guy*

---

[75] Gautier, Axel, Ashwin Ittoo, and Pieter Van Cleynenbreugel. "AI algorithms, price discrimination and collusion: a technological, economic and legal perspective." *European Journal of Law and Economics* 50.3 (2020): 405-435.

[76] Dafoe, Allan, et al. "Open problems in cooperative AI." *arXiv preprint arXiv:2012.08630* (2020).

[77] Schrepel, Thibault. "Collusion by blockchain and smart contracts." *Harvard Journal of Law and Technology* 33 (2019)

[78] https://ec.europa.eu/competition-policy/cartels/leniency_en Accessed on the 13th of July 2021.

[79] Beneke, Francisco, and Mark-Oliver Mackenrodt. "Artificial intelligence and collusion." *IIC-International Review of Intellectual Property and Competition Law* 50.1 (2019): 109-134.

*named Bob to do it, then it probably isn't ok for an algorithm to do it either. [...] So, from my perspective, if conduct was unlawful before, using an algorithm to effectuate it will not magically transform it into lawful behavior. Likewise, using algorithms in ways that do not offend traditional antitrust norms is unlikely to create novel liability scenarios."[80]*. The rest of this section focuses on things algorithms might be able to do, that Bob could not. In other words, we will investigate key properties of algorithms, that have been used to argue for the uniqueness and novelty of the problems posed by algorithmic collusion.

Implicit/tacit collusion

While tacit collusion can be reached in small oligopolies by humans, most papers on algorithmic collusion we reviewed[81] either explicitly mention the specter of algorithmic tacit collusion, or seem to implicitly see it as the main algorithmic threat to antitrust law. This makes sense, as tacit collusion is problematic in all of our three scenarios: While the "Messenger" scenario focuses on algorithms as tools for the stabilization of classical collusive agreements, this does include existing forms of tacit collusion in oligopolies, which could become more easy to arrive at, and potentially harder to detect, if pricing algorithms could in fact increase the efficiency of enforcement. This might be particularly problematic, as competition policy has not been paying a lot of attention to tacit collusion, in part because of its relative rarity so far[82]. While not classically falling under tacit collusion, some versions of the "Hub-and-Spoke" scenario could allow for neither the hub, nor the spokes being aware of the collusive agreement, even though cases, in which the hub consciously facilitates collusion might be more realistic. Lastly, collusion in the archetypical "Independent Agents" scenario is always tacit, as it is the algorithms, not their human deployers, who find a collusive joint strategy, be it with or without the use of explicit communication between them.

Market concentration

The first way in which the use of algorithmic pricing might affect collusion is via increased market concentration. This is not because of any specific properties of algorithms, but because of market dynamics around algorithms, especially AI: Market entries might be limited if pricing algorithms, especially self-learning ones, play an important role in a market, because AI is an active field of research, limiting the amount of firms able to produce AI systems of the best quality, and because many modern machine learning algorithms experience immense returns to scale regarding data[83].

The previous argument seems to be in tension with the broad observation that markets with frequent innovation experience less collusion, because less stable market conditions might increase asymmetries in market conditions while also making more frequent updates to collusive agreements necessary[84]. However, once market conditions become too asymmetric, the likelihood that sustained participation in this market ceases to be profitable for some participants increases. Another mechanism by which innovation can hamper collusion is via the creation of incentives for new firms to enter a market and implement innovations that have been neglected by the incumbents[85]. However, the potential for large initial costs for data acquisition and computational infrastructure, as well as low marginal costs due to network effects and positive feedback loops between algorithm-enabled service quality and algorithm-enabling data availability[86] could limit market entries, especially in markets, where better pricing algorithms convey a strong competitive advantage[87].

---

[80] Ohlhausen, Maureen K. "Should We Fear The Things That Go Beep In the Night? Some Initial Thoughts on the Intersection of Antitrust Law and Algorithmic Pricing." Federal Trade Commission (2017)
https://www.ftc.gov/system/files/documents/public_statements/1220893/ohlhausen_-_concurrences_5-23-17.pdf
[81] See Table 1 in the Appendix
[82] OECD, "Algorithms and Collusion: Competition Policy in the Digital Age" (2017)
www.oecd.org/competition/algorithms-collusion-competition-policy-in-the-digital-age.htm
[83] Kaplan, Jared, et al. "Scaling laws for neural language models." *arXiv preprint arXiv:2001.08361* (2020). and Hernandez, Danny, et al. "Scaling laws for transfer." *arXiv preprint arXiv:2102.01293* (2021).

While the availability of data brokers and third party software developers offering pricing algorithms could seemingly alleviate some of these problems, the problem might just be shifted to another layer in that case. In particular, market concentration at the level of pricing algorithm developers could facilitate "Hub-and-Spoke"-type collusion, in a similar way Uber has been speculated to create an "algorithmic monopoly", coordinating its drivers prices to supracompetitive levels[88].

Speed

The first direct property of algorithms that might mediate algorithms' effects on competition is speed[89] [90]: As argued earlier, quicker punishment of deviations from a collusive strategy can stabilize collusion, and algorithms are undoubtedly able to operate at a higher speed, and adjust prices a lot more frequently than humans. In addition to the previously laid out arguments for how frequent interactions stabilize collusion, the literature on algorithmic collusion has brought forward two additional arguments: Speed can reduce the overall duration and thus cost of price wars, in case they do happen[91] and can similarly reduce the cost of price signals by making them too short for customers, as opposed to other firms that employ pricing algorithms, to react[92]. While speed does not play an important role in the "Hub-and-Spoke" type scenarios, it is one of the primary reasons the "Messenger" scenario might be more problematic than simple human collusion, and can still play an important role in the "Independent Algorithms" scenario, both by increasing the range of stable collusive strategies, and by reducing the overall costs a self-learning algorithm incurs through experimentation while still learning.

---

[84] Bundeskartellamt and Autorité de la concurrence, "Algorithms and Competition" (2019)
[85] Azzutti, Alessio, Wolf-Georg Ringe, and H. Siegfried Stiehl. "Machine Learning, Market Manipulation and Collusion on Capital Markets: Why the Black Box matters." (2021).
[86] OECD Secretariat, "BIG DATA: BRINGING COMPETITION POLICY TO THE DIGITAL ERA --Background note by the Secretariat --" (2016)
[87] Chen, Le, Alan Mislove, and Christo Wilson. "An empirical analysis of algorithmic pricing on amazon marketplace." *Proceedings of the 25th international conference on World Wide Web*. 2016.
[88] Ezrachi, Ariel, and Maurice E. Stucke. "Artificial intelligence & collusion: When computers inhibit competition." *U. Ill. L. Rev.* (2017): 1775.
[89] Mehra, Salil K. "Antitrust and the robo-seller: Competition in the time of algorithms." *Minn. L. Rev.* 100 (2015): 1323.
[90] Ohlhausen, Maureen K. "Should We Fear The Things That Go Beep In the Night? Some Initial Thoughts on the Intersection of Antitrust Law and Algorithmic Pricing." Federal Trade Commission (2017)
[91] Gal, Michal S. "Algorithms as illegal agreements." *Berkeley Tech. LJ* 34 (2019): 67.
[92] OECD, "Algorithms and Collusion: Competition Policy in the Digital Age" (2017)

However, it is worth noting that operating at higher speed reduces the scope for human oversight of pricing algorithms, which while potentially providing a line of legal defence by pleading ignorance of an algorithm's behaviour also comes with significant financial risk, as illustrated by the 2010 Flash Crash[93], United Airlines' algorithm mistakenly selling flight tickets at 5$ a piece[94], and *"The Making of a Fly"* example[95]. The latter, in which an ordinary book was offered for tens of millions of Dollars on the Amazon Marketplace as the result of two interdependent pricing algorithms is frequently cited in support of the plausibility of algorithmic collusion. However, as this price is clearly above what anyone would ever want to pay for this book, the seller almost certainly did not make a profit, and the example rather serves as a warning to firms about the potentially harmful runaway dynamics interdependent algorithms can exhibit.

Efficiency

The next key advantage algorithms have over humans is scalability and efficiency: Computers are able to process large amounts of data, and quickly adjust prices for many different products[96], which would often be prohibitively time consuming and thus expensive for humans to do. This can both increase effective market transparency, as more information about the market can be processed, and the frequency of interactions, as more prices can be updated more frequently. In addition, efficiency might allow for more complex collusive strategies that could be hard to detect, for example based on cross-market punishment. Efficiency is likely to play an important role for all three of our scenarios, but might not be strictly required for "Hub-and-spoke".

Learning capabilities

Self-learning algorithms can learn how to organize patters via unsupervised learning, predict key market events via supervised learning, or learn high-profit strategies via reinforcement learning (RL) algorithms[97] like Q-learning[98]. The first set two types of learning capabilities can play an important role for the "messenger" scenario by creating *de facto* market transparency: Even if key variables like competitors' prices are unobservable, machine learning algorithms could be able to learn to accurately predict them from other variables, such as sales numbers and demand statistics. Similarly, algorithms could enhance a firms "perceptive" capabilities by filtering out the most relevant market information, if data is abundant but difficult to process. These improved capabilities could help firms to better differentiate between defection and "legitimate" responses to demand shifts[99], and reduce

---

noise in firms estimates of market conditions and others' prices[100]. Both of these factors might reduce the propensity for accidental price wars.

However, supervised learning requires potentially large amounts of "ground truth" data to learn how to accurately predict, and such data might not always be available for quantities of interest, such as others' prices. This creates a hen and egg problem, and markets that start out very intransparent might also be the least affected by prediction algorithms: In such markets companies would be forced to rely on unsupervised learning instead, which can still be useful by providing complexity reduction, but does not yield straightforward market predictions.

In addition, there is likely to be a tension between overall prediction quality and robustness/security: While modern deep learning systems strongly outperform other machine learning models in prediction tasks like image classification, they are vulnerable to so-called adversarial examples[101], humanly imperceptible modifications to the input data that still lead to big changes in the prediction output. Obviously, wrong predictions about market conditions, if believed by humans or directly connected to a pricing algorithm can have large negative consequences, as prices could be set in a very disadvantageous manner. As the "adversarial" in adversarial examples suggests, the basic threat model is that such examples could be created by an adversary to manipulate a deep learning systems' behaviour to their advantage. As there are many other firms, i.e. potential adversaries in a market, and as these might have explicit control over some of the input data a firms' algorithm receives[102], firms might not necessarily prefer deep learning models over simpler models that are more robust at the cost of overall prediction capabilities. This is especially true, since the issue with adversarial examples is indicative of a larger problem with the robustness of deep neural networks[103]. Accordingly, recent progress in deep learning[104] does not directly transfer to the business context, especially in competitive contexts with multiple adversaries amd incentive to exploit each other. In addition, data on customer purchases might be a lot easier to obtain, and improved prediction capabilities allow for better estimations of customers' willingness to pay. This incentivizes firms to employ personalized pricing and more generally better segment markets, which can make collusion more difficult to sustain[105], at least outside the "Hub-and-Spoke" scenario[106], both by making prices even harder to observe for competitors, and by increasing the complexity of collusive agreements, that now need to specify multiple collusive prices for different segments instead of a single price.

---

[100] Mehra, Salil K. "Antitrust and the robo-seller: Competition in the time of algorithms." *Minn. L. Rev.* 100 (2015): 1323.

[101] Goodfellow, Ian J., Jonathon Shlens, and Christian Szegedy. "Explaining and harnessing adversarial examples." *arXiv preprint arXiv:1412.6572* (2014).

[102] In particular, they control their own market activities, and might even be able to falsify information about some activities that play into another firm's prediction algorithm.

[103] Zheng, Stephan, et al. "Improving the robustness of deep neural networks via stability training." *Proceedings of the ieee conference on computer vision and pattern recognition*. 2016.

[104] Hernandez, Danny, and Tom B. Brown. "Measuring the algorithmic efficiency of neural networks." *arXiv preprint arXiv:2005.04305* (2020).

[105] Beneke, Francisco, and Mark-Oliver Mackenrodt. "Remedies for algorithmic tacit collusion." *Journal of Antitrust Enforcement* 9.1 (2021): 152-176.

[106] Mehra, Salil K. "Price Discrimination-Driven Algorithmic Collusion: Platforms for Durable Cartels." *Stan. JL Bus. & Fin.* 26 (2021): 171.

While learning capabilities might not be crucial in the "Hub-and-Spoke" scenario, where the main point is not the algorithms' novel capabilities, but the fact that the use of algorithms provides structure for a hub-and-spoke cartel, learning capabilities are obviously crucial in the "Independent Algorithms" scenario that is centered around self-learning algorithms.

However, it is unlikely that algorithms are already sufficiently good at learning strategies to have an immediate impact on algorithmic collusion: While there have been many impressive examples of algorithms learning complex strategies in various simulations in the recent past, including deep reinforcement learning systems beating the human world champion in the game of Go[107] or learning to cooperate in teams and showing complex coordinated strategies in a simulated catching game[108], these systems usually take very long times to learn. For example, Agent57, DeepMind's recent system that is able to play many Atari games on a superhuman level[109] required over 40 years of training data[110], which was only possible due to massively sped up and highly parallelized simulations of the respective games. Obviously, slow learning and bad initial performance is highly problematic from a commercial point of view, especially if the system has to be operating in a real market to learn[111]. As reinforcement learning often relies on the algorithm's ability to experiment and try out new behaviour to observe its effects, operation in a real market might be necessary. While "offline" forms of reinforcement learning that solely rely on passively observed data exist and have been suggested as a "solution" by Ezrachi and Stucke[112], these approaches do lag behind in terms of performance[113], and are likely unfit for multi-agent interactions such as markets, as the passively collected data cannot take into account the shifting strategy of other firms. A similar problem goes for first training in a simulation, where the algorithm can experiment. Indeed, it has been shown that even slight differences in the strategies two deep reinforcement learning agents faced while learning to play multiplayer games can have large negative effects on their ability to coordinate, unless complicated prevention measures are taken[114]. While it might theoretically be possible to accurately simulate both shifting market conditions and the strategic responses of other firms to an algorithms experimental actions, such capabilities are far beyond, and could potentially eliminate the need for antitrust law altogether, as a centralized algorithmic economy might very well outperform markets with such advanced computational and algorithmic capabilities[115].

---

[107] Silver, David, et al. "Mastering the game of go without human knowledge." *nature* 550.7676 (2017): 354-359.
[108] Baker, Bowen, et al. "Emergent tool use from multi-agent autocurricula." *arXiv preprint arXiv:1909.07528* (2019).
[109] Badia, Adrià Puigdomènech, et al. "Agent57: Outperforming the atari human benchmark." *International Conference on Machine Learning*. PMLR, 2020.
[110] Dorner, Florian E. "Measuring Progress in Deep Reinforcement Learning Sample Efficiency." *arXiv preprint arXiv:2102.04881* (2021).
[111] Bundeskartellamt and Autorité de la concurrence, "Algorithms and Competition" (2019)
[112] Ezrachi, Ariel, and Maurice E. Stucke. "Sustainable and unchallenged algorithmic tacit collusion." *Nw. J. Tech. & Intell. Prop.* 17 (2019): 217.
[113] Levine, Sergey, et al. "Offline reinforcement learning: Tutorial, review, and perspectives on open problems." *arXiv preprint arXiv:2005.01643* (2020).
[114] Lanctot, Marc, et al. "A unified game-theoretic approach to multiagent reinforcement learning." *arXiv preprint arXiv:1711.00832* (2017).
[115] Parson, Edward Ted A. "Max–A Thought Experiment: Could AI Run the Economy Better Than Markets?." (2020)

In addition, both many deep reinforcement learning algorithms[116] and multi-agent reinforcement learning[117] lack theoretical performance guarantee, and this dearth of theoretical guarantees is very likely to be even worse when both are combined. Similarly, robustness concerns are even more severe for independently acting self-learning agents, as shown by the existence of adversarial policies[118], seemingly random behavioural patterns that can be used to make a usually very component adversary in a physics-engine based sports simulation flop around uselessly. Lastly, it is worth noting that most of the recent milestones in multi-agent reinforcement learning have been in zero-sum games like GO[119], rather than cooperative games that require coordination, or mixed motive scenarios like markets that require both coordination and commitment or punishment.

So, while the renewed research interest in mixed-motive games[120] and the fast trends in the data efficiency of deep reinforcement learning[121] systems suggest, that algorithmic collusion by deep reinforcement learning agents might very well become possible at some point in the future, this seems very unlikely to happen within the next few years. While simpler algorithms might learn faster, circumvent some of deep reinforcement learning's robustness issues, and have been shown to be able to implement collusive strategies in simple markets[122], these algorithms are not able to use the large amounts of available data and thus do not benefit from the improvements in prediction quality[123]. Furthermore, complex algorithms might very well be need to deal with the complexities[124] of real-world markets[125]. In addition, most of these simple algorithms have been around for years or decades[126], without any reported cases of algorithmic collusion by such algorithms[127], indicating that they do not pose a threat to antitrust either.

### Behavioural consistency

Algorithms might be able to consistently stick to a strategy better than human, and are not influenced by human whims and emotions like anger[128] that could trigger unnecessary price wars. In addition, algorithms might not be motivated by emotions like fear of detection to back out of collusive agreements[129]. However, unless there is a liability gap for algorithms, firms might actually want their algorithms to avoid detection[130], which can require the use of more complicated algorithms[131], such that human emotions might be a feature, not a bug[132]. Similarly, human intuitions around the badness of cheating could stabilize collusion[133]. In addition, the consistent execution of a strategy might not actually be desirable, as the previous discussion of the grim trigger strategy illustrates. Similarly, human "inconsistency" rather than algorithmic consistency could plausibly have prevented both the 2010 Flash Crash[134] and the "The Making of a Fly" case[135].

---

[116] Mnih, Volodymyr, et al. "Playing atari with deep reinforcement learning." *arXiv preprint arXiv:1312.5602* (2013).
[117] Bowling, Michael. "Convergence problems of general-sum multiagent reinforcement learning." *ICML*. 2000.
[118] Gleave, Adam, et al. "Adversarial policies: Attacking deep reinforcement learning." *arXiv preprint arXiv:1905.10615* (2019).
[119] Dafoe, Allan, et al. "Open problems in cooperative AI." *arXiv preprint arXiv:2012.08630* (2020).
[120] Id.
[121] Dorner, Florian E. "Measuring Progress in Deep Reinforcement Learning Sample Efficiency." *arXiv preprint arXiv:2102.04881* (2021).
[122] Brown, Zach Y., and Alexander MacKay. *Competition in pricing algorithms*. No. w28860. National Bureau of Economic Research, 2021.
[123] While deep reinforcement learning algorithms do not necessarily directly employ supervised prediction algorithms as a subroutine, they do make use of neural networks that are structured very similar to supervised learning systems to effectively process information
[124] Deng, Ai. "What Do We Know about Algorithmic Tacit Collusion?." (2018).
[125] Schwalbe, Ulrich. "Algorithms, machine learning, and collusion." *Journal of Competition Law & Economics* 14.4 (2018): 568-607.
[126] Watkins, Christopher JCH, and Peter Dayan. "Q-learning." *Machine learning* 8.3-4 (1992): 279-292.
[127] Such a case would be very unlikely to have escaped all of the authors warning about algorithmic collusion's attention
[128] Beneke, Francisco, and Mark-Oliver Mackenrodt. "Remedies for algorithmic tacit collusion." *Journal of Antitrust Enforcement* 9.1 (2021): 152-176.

Other arguments we found in the literature include algorithms' supposed freedom of human biases[136] [137], which would at least require further details given the vast literature on various forms of bias in machine learning[138], including the theoretical importance of inductive bias for learning[139], as well as the reduction of principal-agent problems: Human employees might be motivated by short term gains linked to promotions or boni to undercut prices, regardless of long term consequences for the firm[140] [141]. However, similar problems have been identified for learning algorithms[142], even though there clearly exist relevant differences in the extent to which firms can alter humans' vs algorithms' incentives.

Again, the collected arguments are most relevant for the "Messenger" and "Independent Algorithm" scenarios, but seem rather weak overall, especially concerning the second scenario involving learning agents.

---

## (Algorithmic) Transparency

Algorithmic transparency refers to the fact that algorithms consist of code that can in principle be read by humans and perhaps even other algorithms to understand an algorithms' exact decision procedure. Note, that this maps to the second type of market transparency we discussed earlier, while algorithms' effects on the first are discussed were discussed in earlier subsections. Many arguments for algorithmic collusion seem to strongly rely on the assumption, that firms or algorithms can "decode" each others algorithms[143] [144] [145]. While this might be possible for simple pricing algorithms, modern neural networks which might be required to handle the complexity of real-world markets are notoriously hard to understand. Furthermore, while transparent algorithms could be used to communicate a firm's strategy, this form of communication is still relatively limited and potentially unfit for tasks beyond coordination on a symmetric collusive equilibrium. In addition, courts could potentially interpret excessive transparency that clearly facilitates collusion as evidence of a collusive agreement. Lastly, while transparency could help algorithms learn to collude, this is far from trivial: While learning how to respond to a fixed transparent algorithm should be easy, two learning algorithms might need to somehow get around a problem of infinite recursion: If one algorithm's behaviour depends on its beliefs about the other algorithm, which in turn include beliefs about the first algorithm's behaviour, there might not always be a stable solution. While there might be ways to work around this problem[146], we are not aware of any work that successfully uses algorithmic transparency to facilitate coordination in complex environments.

There are also multiple severe downsides to increased transparency for firms: The more clearly an algorithm is designed to issue punishment for deviations from a collusive price, the more likely it seems that courts would see it as sufficient evidence for an illegal agreement to collude. In addition, the previously discussed exploits of deep neural networks via adversarial examples and policies are usually easier to execute, the more information the adversary has about the system[147]. In combination with incentives to protect intellectual property both in the form of trained neural network weights, and more broadly algorithmic innovations, it is far from clear that firms would opt to provide extensive information about their pricing algorithms. In fact, many observers have worried about intransparency around algorithmic collusion[148] [149], especially around neural networks[150] [151], rather than transparency: The harder it is to understand what an algorithm is doing for both the deploying firm and the antitrust authorities, the harder it becomes to prove the firms' awareness of its algorithm's anticompetitive behaviour, or even that behaviour's anticompetitiveness itself.

---

[143] Gal, Michal S. "Algorithms as illegal agreements." *Berkeley Tech. LJ* 34 (2019): 67.

[144] Ezrachi, Ariel, and Maurice E. Stucke. "Sustainable and unchallenged algorithmic tacit collusion." *Nw. J. Tech. & Intell. Prop.* 17 (2019): 217.

[145] Salcedo, Bruno. "Pricing Algorithms and Tacit Collusion." *Manuscript, Pennsylvania State University* (2016).

[146] LaVictoire, Patrick, et al. "Program Equilibrium in the Prisoner's Dilemma via Löb's Theorem." *Workshops at the Twenty-Eighth AAAI Conference on Artificial Intelligence*. 2014.

[147] Ilyas, Andrew, et al. "Black-box adversarial attacks with limited queries and information." *International Conference on Machine Learning*. PMLR, 2018.

[148] Ezrachi, Ariel, and Maurice E. Stucke. "Sustainable and unchallenged algorithmic tacit collusion." *Nw. J. Tech. & Intell. Prop.* 17 (2019): 217.

[149] Monopolkommission, "Wettbewerb 2018 XXII. Hauptgutachten der Monopolkommission gemäß §44 Abs.1 Satz 1 GWB" (2018) https://monopolkommission.de/images/HG22/HGXXII_Gesamt.pdf

[150] Azzutti, Alessio, Wolf-Georg Ringe, and H. Siegfried Stiehl. "Machine Learning, Market Manipulation and Collusion on Capital Markets: Why the Black Box matters." (2021).

[151] Beneke, Francisco, and Mark-Oliver Mackenrodt. "Remedies for algorithmic tacit collusion." *Journal of Antitrust Enforcement* 9.1 (2021): 152-176.

Once again, there is a split between the "Hub-and-Spoke" scenario and the other two: In the former, algorithmic transparency might be strictly detrimental to collusion, as it increases the likelihood that participants could become aware of the extent of collusion and potentially liable. In the other scenarios, both the direction of the effect of increased transparency on collusion, and the relative strength of incentives for firms to aim for or avoid transparency are unclear.

## Communication and Coordination

While some authors focus on the potential for algorithms to learn to effectively coordinate[152] [153] [154], perhaps as other factors like speed simplify collusion and make communication less important[155] [156], others point out the lack of principled arguments supporting this potential[157], and that current reinforcement learning algorithms' capabilities to coordinate in complex environments or learn to communicate are limited at best[158].

One way in which algorithms might communicate is via the use of price signals[159] [160], but the utility of price signals might be limited, especially in asymmetric markets, as discussed earlier. While communication could be hard-coded into human designed algorithms in the "Messenger" scenario, this would likely be interpreted as strong evidence for collusive intent. In the "Independent Algorithms" scenario, one key challenge around communication is the establishment of common ground for algorithms to base their communication on[161]. This might be particularly difficult, because such a grounding and common understanding might need to be bootstrapped and because the mixed-motive nature of market interactions could incentivize algorithms to "lie" during early phases of "language development", which in turn might reduce the incentive for other algorithms to "listen". Human language development circumvents this problem by humans learning to communicate from their parents, who a) already speak a fully developed language and b) have limited incentive to manipulate their child. In the "Hub-and-Spoke" scenario, coordination is covered by the hub, such that communication is only required between the hub and the spokes. Assuming that the hub's earning are a percentage of the spokes', incentives are also aligned such that the spokes should communicate accurate information to the hub, which in turn provides them with a profit maximizing price.

---

[152] OECD, "Algorithms and Collusion: Competition Policy in the Digital Age" (2017) www.oecd.org/competition/algorithms-collusion-competition-policy-in-the-digital-age.htm
[153] Bernhardt, Lea, and Ralf Dewenter. "Collusion by code or algorithmic collusion? When pricing algorithms take over." *European Competition Journal* 16.2-3 (2020): 312-342.
[154] Ballard, Dylan I., and Amar S. Naik. "Algorithms, artificial intelligence, and joint conduct." *Antitrust Chronicle* 2 (2017): 29.
[155] Ezrachi, Ariel, and Maurice E. Stucke. "Sustainable and unchallenged algorithmic tacit collusion." *Nw. J. Tech. & Intell. Prop.* 17 (2019): 217.
[156] Mehra, Salil K. "Antitrust and the robo-seller: Competition in the time of algorithms." *Minn. L. Rev.* 100 (2015): 1323.
[157] Kühn, Kai-Uwe, and Steve Tadelis. "Algorithmic collusion." *Presentation prepared for CRESSE, URL https://www.ebos.com.cy/cresse2013/uploadfiles/2017_sps5_pr2.pdf* (2017).
[158] Schwalbe, Ulrich. "Algorithms, machine learning, and collusion." *Journal of Competition Law & Economics* 14.4 (2018): 568-607.
[159] Ohlhausen, Maureen K. "Should We Fear The Things That Go Beep In the Night? Some Initial Thoughts on the Intersection of Antitrust Law and Algorithmic Pricing." Federal Trade Commission (2017)
[160] Monopolkommission, "Wettbewerb 2018 XXII. Hauptgutachten der Monopolkommission gemäß §44 Abs.1 Satz 1 GWB" (2018)
[161] Dafoe, Allan, et al. "Open problems in cooperative AI." *arXiv preprint arXiv:2012.08630* (2020).

Commitment

If it is sufficiently costly or slow for firms to change the pricing algorithms they use, pricing algorithms could essentially act as a form of commitment device supporting the credibility of threats involving costly forms of punishment[162] [163]. In addition, it has been argued that the fast speed algorithms operate on could help to make commitment credible[164]: Punishment in the form of undercutting prices might be less costly for a firm, if it is only sustained shortly. While this decreases the deterrence effect of the punishment, it could still be effective if it causes other firms' costs for deviations to exceed the benefits. And repeated demonstrations of such punishment, which are now cheaper, should increase others' certainty, that the punishment will be applied consistently.

Commitment plays a subordinate role in the "Hub-and-Spoke" scenario, as spokes can only unilaterally deviate from the agreement by deciding to switch their pricing algorithm. In addition, unilateral agreements between the hub and individual spokes might be legally enforceable, which might reduce spokes' incentives to change algorithms, for example if they could not do so without having to keep paying royalties to the initial algorithm provider. As the arguments from the literature involve algorithms as a commitment device, rather than learning algorithms commiting to a particular strategy, it seems likely that commitment is going to be play a more important role in the "Messenger" than in the "Independent Algorithms" scenario. It is also worth highlighting, that algorithms could enable commitment in a way that is not captured in any of the scenarios, as it does not necessarily involve pricing algorithms: The blockchain and smart contracts could facilitate collusive agreements that courts would not hold up[165] by tying compliance to cryptocurrency payments. Such agreements might also be hard to prove for courts, if they are based on a private blockchain[166].

Symmetry

Symmetry between pricing algorithms is often cited as a reason for the plausibility of algorithmic collusion[167] [168] [169]. While it indeed sounds intuitive that two firms using the same (learning) algorithm might increase the likelihood of algorithmic collusion, this intuition does not seem to be backed up in detail by those warning about the dangers of algorithmic collusion. Obviously, some strategies might facilitate collusion with firms employing the same strategy, but collusive outcomes can easily be achieved with asymmetric strategies[170], and many symmetric strategies do not involve any collusion. While it seems plausible, that two copies of the same algorithm will reach *an* equilibrium more easily, due to the existence of more obvious Schelling points[171], it is not entirely clear why symmetry (on its own) could favour more collusive equilibria. Ezrachi and Stucke claim that it might be easier for similar computers to anticipate and understand moves by other machines[172], but this claim is not elaborated further. Perhaps, the use of similar input variables could prevent algorithm A from accidentally starting a price war by reacting to a change in a variable that B does not observe, which is then interpreted and punished by B as a defection. Alternatively, symmetry might help with bootstrapping communication by providing common grounding, and similar learning speeds between two symmetric algorithms could prevent one of them from becoming sufficiently dominant to stop the other one from learning[173]. Lastly, symmetric training in the same market simulation might be able to alleviate some of the obstacles to training learning algorithms without having to deploy them during early learning[174].

---

[162] Ezrachi, Ariel, and Maurice E. Stucke. "Sustainable and unchallenged algorithmic tacit collusion." *Nw. J. Tech. & Intell. Prop.* 17 (2019): 217.
[163] Salcedo, Bruno. "Pricing Algorithms and Tacit Collusion." *Manuscript, Pennsylvania State University* (2016).
[164] Mehra, Salil K. "Antitrust and the robo-seller: Competition in the time of algorithms." *Minn. L. Rev.* 100 (2015): 1323.
[165] Schrepel, Thibault. "Collusion by blockchain and smart contracts." *Harvard Journal of Law and Technology* 33 (2019)
[166] Id.
[167] Ezrachi, Ariel, and Maurice E. Stucke. "Artificial intelligence & collusion: When computers inhibit competition." *U. Ill. L. Rev.* (2017): 1775.
[168] OECD, "Algorithms and Collusion: Competition Policy in the Digital Age" (2017) www.oecd.org/competition/algorithms-collusion-competition-policy-in-the-digital-age.htm
[169] Bernhardt, Lea, and Ralf Dewenter. "Collusion by code or algorithmic collusion? When pricing algorithms take over." *European Competition Journal* 16.2-3 (2020): 312-342.

However, both the likely symmetry of algorithms[175] [176] and the symmetry of firms' deploying firms' market conditions[177] have been called into question[178] repeatedly. The second matters, as many of the arguments about symmetric algorithms implicitly require symmetric tasks for these algorithms. For example, the same algorithm might learn a lot more quickly if deployed by a firm that controls most of the market, rather than a smaller firm, whose actions might only have small effects on the overall market[179]. Firms in most markets are not uniform, and feature different production costs, distribution channels[180], data sources[181], and technical capabilities or budget for third party software developers. In addition, many modern machine learning algorithms employ stochastic training[182], which is likely to break initial symmetry.

---

[170] Such as "tit-for-tat" and "always cooperate" in an iterated prisoner's dilemma.

[171] Even that is not a given, as can be seen by looking at two agents that update simultaneously to their respective best response in a game that rewards them for picking actions that are different from each other.

[172] Ezrachi, Ariel, and Maurice E. Stucke. "Artificial intelligence & collusion: When computers inhibit competition." *U. Ill. L. Rev.* (2017): 1775.

[173] Similar to how humans learn best, by doing tasks that are at the boundary of their capabilities, neural networks might learn very inefficiently when presented with tasks that are too easy or too hard, such as maximizing profit against a much more capable pricing algorithm. Bengio, Yoshua, et al. "Curriculum learning." *Proceedings of the 26th annual international conference on machine learning*. 2009.

[174] Lanctot, Marc, et al. "A unified game-theoretic approach to multiagent reinforcement learning." *arXiv preprint arXiv:1711.00832* (2017).

[175] Deng, Ai. "What Do We Know about Algorithmic Tacit Collusion?." (2018).

[176] Petit, Nicolas. "Antitrust and artificial intelligence: a research agenda." *Journal of European Competition Law & Practice* 8.6 (2017): 361-362.

[177] Gal, Michal S. "Algorithms as illegal agreements." *Berkeley Tech. LJ* 34 (2019): 67.

[178] Bundeskartellamt and Autorité de la concurrence, "Algorithms and Competition" (2019)

[179] In the latter case, it could be harder for the algorithm to infer the causal effect of its actions on the market, as the effect is smaller, and other firms' actions introduce relatively larger noise.

[180] Beneke, Francisco, and Mark-Oliver Mackenrodt. "Remedies for algorithmic tacit collusion." *Journal of Antitrust Enforcement* 9.1 (2021): 152-176.

[181] Gautier, Axel, Ashwin Ittoo, and Pieter Van Cleynenbreugel. "AI algorithms, price discrimination and collusion: a technological, economic and legal perspective." *European Journal of Law and Economics* 50.3 (2020): 405-435.

[182] Id.

Symmetry plays a similar role in the "Messenger" and "Hub-and-Spoke" scenarios, where firms or a central software developer might deploy symmetric algorithms, perhaps specifically designed for collusion. In the "Independent Algorithms" scenario, symmetry might also be a conscious choice by firms wanting to collude, as they might belief that symmetry enhances cooperation. Alternatively, firms might plausibly symmetrically deploy versions of the same open-source implementation of some reinforcement learning algorithm[183], perhaps including similar industry-specific adaptions.

## The prevalence of pricing algorithms and their use for collusion

There is ample evidence for the adoption of pricing algorithms, especially in online markets: a 2015-2017 EU survey of online retailers has found two-thirds of them to use pricing algorithms[184], while another 2016 study found hundreds of sellers in the Amazon Marketplace showing strong indications of algorithmic pricing[185]. Also, Amazon itself saw a tenfold surge in the amount of daily prices changes between December 2012 and 2013[186], which is strongly indicative of an adoption of pricing algorithms in that time frame. Pricing algorithms also play a role in at least some offline markets, and a recent study found evidence of algorithmic pricing in 20% of a sample roughly 16000 german gas stations[187], as well as associated margin rises, especially for local duopolies. Ezrachi and Stuck cite promotional material by a provider of pricing algorithms promoting them as a means to avoid price wars and describe a case in which roughly 25% of the Danish retail fuel market switched to a pricing algorithm and subsequently increased margins by 5%[188]. Lastly, an OECD report describes a confirmed case of algorithmic collusion from 2015: Several sellers on the US Amazon marketplace had conspired to use pricing algorithms implementing a collusive agreement[189]. A similar case was reported by the british Competition and Markets authority (CMA)[190]. Overall, pricing algorithms seem to have been fairly common, even 5 years ago, and it would be if their use did not continue grow. There also have been multiple cases of algorithms being used for collusion, and it would hardly be surprising, if more and more of these were to be uncovered in the future.

---

[183] For example using OpenAI's baselines https://github.com/openai/baselines
[184] European Commision, "REPORT FROM THE COMMISSION TO THE COUNCIL AND THE EUROPEAN PARLIAMENT, Final report on the E-commerce Sector Inquiry" (2017) http://ec.europa.eu/competition/antitrust/sector_inquiry_final_report_en.pdf accessed on the 14th of July 2021
[185] Chen, Le, Alan Mislove, and Christo Wilson. "An empirical analysis of algorithmic pricing on amazon marketplace." *Proceedings of the 25th international conference on World Wide Web*. 2016.
[186] Competition and Markets Authority. "Pricing algorithms; Economic working paper on the use of algorithms to facilitate collusion and personalised pricing" (2018) https://assets.publishing.service.gov.uk/government/uploads/system/uploads/attachment_data/file/746353/Algorithms_econ_report.pdf
[187] Assad, Stephanie, et al. "Algorithmic pricing and competition: Empirical evidence from the German retail gasoline market." (2020).
[188] Ezrachi, Ariel, and Maurice E. Stucke. "Sustainable and unchallenged algorithmic tacit collusion." *Nw. J. Tech. & Intell. Prop.* 17 (2019): 217.
[189] OECD, "Algorithms and Collusion: Competition Policy in the Digital Age" (2017) www.oecd.org/competition/algorithms-collusion-competition-policy-in-the-digital-age.htm
[190] Competition and Markets Authority. "Pricing algorithms; Economic working paper on the use of algorithms to facilitate collusion and personalised pricing" (2018)

It is however not really clear whether firms are employing learning algorithms, rather than static human-designed algorithms as their pricing algorithms, perhaps except for when they are working with personalized prices. But, as laid out earlier, personalized pricing and algorithmic collusion might not work in concert. Furthermore, the only claim of reinforcement learning being used for financial real world tasks is only backed up by a proof-of-concept paper and not information on real-world usage[191]. However, the british CMA has found providers of pricing algorithms that advertise their use of machine learning[192]. Still, our literature review did not identify any cases of collusion between learning algorithms, and in fact many papers on algorithmic collusion cite the lack of direct empirical evidence[193] [194]. While this is sometimes interpreted as evidence for the difficulty of identifying such collusion[195], it is even stronger evidence that such collusion is not currently happening in complex real-world markets, as also suggested by the large amount of obstacles for algorithms to learn to collude in such markets we identified in the last section.

## What do modelling results tell us?

In addition to theoretical arguments and real-world evidence, the literature on algorithmic collusion contains various papers exploring the effect certain aspects of algorithmic pricing might have in game theoretical models and/or training reinforcement learning algorithms[196] in such models. As noted by Schwalbe[197], these models often do not come close to matching the complexity of real-world markets and assume symmetry in both market conditions and employed algorithms, both of which might be unrealistic. In addition, the impact of algorithms is often modelled quite simplistically, and takeaways can be mutually exclusive due to subtle differences in the model. For example, a certain form of asymmetry facilitates collusion in one example[198], while symmetry is crucial for collusion in another[199]. Regarding reinforcement learning, authors regularly find algorithms achieving supracompetitive prices[200] [201], but a) these prices are often lower than the optimal sustainable collusive price for no apparent reason and b) explanations for why humans would not be able to tacitly collude in the same simple market simulation are often lacking. Because of the simplicity of these modelling results, and the other issues laid out, the relevance of the results might be somewhat limited, and we won't discuss them in further detail in the main text[202].

---

[191] Azzutti, Alessio, Wolf-Georg Ringe, and H. Siegfried Stiehl. "Machine Learning, Market Manipulation and Collusion on Capital Markets: Why the Black Box matters." (2021).

[192] Competition and Markets Authority. "Pricing algorithms; Economic working paper on the use of algorithms to facilitate collusion and personalised pricing" (2018)

[193] Harrington, Joseph E. "Developing competition law for collusion by autonomous artificial agents." *Journal of Competition Law & Economics* 14.3 (2018): 331-363.

[194] Lee, Kenji. "Algorithmic Collusion & Its Implications for Competition Law and Policy." *Available at SSRN 3213296* (2018).

[195] OECD, "Algorithms and Collusion: Competition Policy in the Digital Age" (2017) www.oecd.org/competition/algorithms-collusion-competition-policy-in-the-digital-age.htm

[196] Mostly Q-learning Watkins, Christopher JCH, and Peter Dayan. "Q-learning." *Machine learning* 8.3-4 (1992): 279-292.

[197] Schwalbe, Ulrich. "Algorithms, machine learning, and collusion." *Journal of Competition Law & Economics* 14.4 (2018): 568-607.

[198] Brown, Zach Y., and Alexander MacKay. *Competition in pricing algorithms*. No. w28860. National Bureau of Economic Research, 2021.

[199] Hettich, Matthias. "Algorithmic Collusion: Insights from Deep Learning." *Available at SSRN 3785966* (2021).

[200] Calvano, Emilio, et al. "Artificial intelligence, algorithmic pricing, and collusion." *American Economic Review* 110.10 (2020): 3267-97.

[201] Klein, Timo. "Autonomous algorithmic collusion: Q-learning under sequential pricing." *Amsterdam Law School Research Paper* 2018-15 (2019): 2018-05.

# Recommendations for antitrust policy

## Messenger scenario

As the messenger scenario is concerned with the use of algorithms as mere tools to facilitate existing forms of collusion, one might at first sight be tempted to argue that the scenario cannot be important from an antitrust law perspective, as it does not involve anything fundamentally novel. However, current antitrust law has two flaws that reduce its current effectiveness in preventing societally harmful forms of collusion: First, some forms of illegal collusion might be hard to identify for antitrust authorities[203]. Second, collusion might be legal, unless it involves an agreement to collude or at least anticompetitive intent. Both of these flaws might be exacerbated by the use of pricing algorithms: Pricing algorithms could implement pricing patterns that do not look collusive at first sight, and the efficiency of algorithms reduces the number of humans needed to implement a collusive strategy, which can reduce the chance of one of them reporting the illegal behaviour to the authorities. In addition, the increased speed of pricing algorithms compared to humans makes punishment strategies more effective, and could enable tacit forms of collusion in markets that previously were not susceptible to these.

While short-term rational, non-costly punishment[204] might already be enough to stabilize collusion if administered swiftly, it seems infeasible to forbid firms from reacting rationally to changing market conditions. So while pricing algorithms might add fuel to the debate on whether classic oligopoly behaviour should be illegal[205], preventing such behaviour could require broader reforms of antitrust legislation, and it seems prudent to not attempt these without strong quantitative evidence on whether the damages caused by such behaviour and the additional damages that might be caused by algorithmic pricing outweigh the potential costs of administrative overhead from forcing firms to adopt certain prices, as we well as a potential slowdown regarding innovations around pricing algorithms. The latter point also applies to restrictions on the frequency of price changes by pricing algorithms[206], which might reduce algorithms' effect on the feasibility of collusion at the cost of limiting efficiency improvements by pricing algorithms, and a potential extension to merger review to the creation of larger oligopolies rather than duo- and monopolies[207].

On the other hand, costly punishment, which is more effective at establishing collusion should be interpreted as strong evidence of anticompetitive intent: Its only purpose is to inflict damage on all involved firms in response to deviations from a collusive price, in order to disincentivize others from deviating. Luckily, pricing algorithms could make it easier to establish costly punishment: Unlike with humans, an algorithms' full strategy can in principle be observed, not just its behaviour[208]. In particular, for the comparatively simple, human-designed algorithms we are concerned with here, this can be done by reading and understanding their code[209]. This way, legal defenses involving feigned procompetitive motives for particular instances of costly punishment would cease to work. In addition, different functionalities of simple algorithms might be analyzed in isolation[210], potentially preventing firms from justifying an algorithm's punishment strategy with other procompetitive functionalities.

---

[202] Further details on the models and other experimental details for all results we reviewed can be found in Table 1 of the Appendix.
[203] Which might explain the popularity of leniency programs as an antitrust tools. Schrepel, Thibault. "Collusion by blockchain and smart contracts." *Harvard Journal of Law and Technology* 33 (2019)
[204] That is thus a byproduct of competitive behaviour
[205] OECD, "Algorithms and Collusion: Competition Policy in the Digital Age" (2017) www.oecd.org/competition/algorithms-collusion-competition-policy-in-the-digital-age.htm
[206] Ezrachi, Ariel, and Maurice E. Stucke. "Artificial intelligence & collusion: When computers inhibit competition." *U. Ill. L. Rev.* (2017): 1775.
[207] Id.

Additionally, an adjustment of a firm's pricing algorithm that avoids costly punishment by another firm's algorithm after this punishment has been observed repeatedly, and has no apparent other purpose, might reasonably be interpreted as a "meeting of minds", such that firms could even potentially be prosecuted under TFEU 101 or the Sherman act. As argued by Joseph Harrington, the potential ability to prove agreements without the need for explicit communication could be a step towards effectively rendering collusion, rather than the communication facilitating it illegal[211].

To that extent, antitrust authorities should make sure to have sufficient in-house expertise on pricing algorithms, both from a technical and a business point of view, to both be able to quickly identify markers of algorithmic collusion, both in code and in observed market dynamics, and make a convincing case for why a given algorithm used to facilitate collusion provides evidence for illegal conduct. Prevention of algorithmic collusion could also be aided by requiring firms to report their use of pricing algorithms[212], and perhaps even implementation details, to antitrust authorities. Furthermore, preventing firms from obfuscating their algorithms' decision making process could be prevented by transparency requirements[213], combined with prohibitions to share these transparent algorithms with other firms. Instead of strict transparency requirements, a reversal of the burden of proof around seemingly collusive behaviour could be implemented for firms that use intransparent algorithms, but this might require larger changes to existing antitrust law[214], and might again hamper innovation.

## Hub-and-Spoke scenario

Regarding algorithmic hub-and-spoke collusion, two big issues are not about algorithmic collusion per se, but rather about market power: As laid out earlier, concentration around a few or a single developer of pricing algorithms might emerge because of economies of scale. This way, Hub-and-Spoke collusion might become more likely just because an usually large amount of pricing decisions are, at least indirectly, controlled by a few actors. In addition, the use of pricing algorithms might enable the hub to implement more complicated collusive strategies, including price discrimination[215] which could drastically reduce the negative impact of market concentration on consumer welfare.

---

[208] Harrington, Joseph E. "Developing competition law for collusion by autonomous artificial agents." *Journal of Competition Law & Economics* 14.3 (2018): 331-363.

[209] If the algorithm is more complex, perhaps as it is used by a large firm producing and selling many different products, this approach can be supplemented with queries for counterfactual inputs that allow some insight into how different variables affect the pricing decisions.

[210] Gal, Michal S. "Algorithms as illegal agreements." *Berkeley Tech. LJ* 34 (2019): 67.

[211] Harrington, Joseph E. "Developing competition law for collusion by autonomous artificial agents." *Journal of Competition Law & Economics* 14.3 (2018): 331-363.

[212] Ezrachi, Ariel, and Maurice E. Stucke. "Artificial intelligence & collusion: When computers inhibit competition." *U. Ill. L. Rev.* (2017): 1775.

[213] Bernhardt, Lea, and Ralf Dewenter. "Collusion by code or algorithmic collusion? When pricing algorithms take over." *European Competition Journal* 16.2-3 (2020): 312-342.

[214] Monopolkommission, "Wettbewerb 2018 XXII. Hauptgutachten der Monopolkommission gemäß §44 Abs.1 Satz 1 GWB" (2018)

The automatic control of pricing decisions also eliminates the need for explicit communication in maintaining a classical hub-and-spoke cartel. While algorithmic hub-and-spoke collusion without the hub's knowledge seems unlikely, unless the hub is using self-learning algorithms to maximize joint profit[216], unawareness by the spokes is entirely plausible. In fact, the spokes gain very little from knowledge about the hub's collusive activities, so they are likely to be unaware, unless they actively chose the hub because of knowledge about its collusive activities. But if the spokes are not aware of the collusion, there is no agreement between them and the hub, excluding all involved parties from liability under TFEU, Article 101[217]. While the spokes might still be liable under TFEU, Article 101, if they "could reasonably have forseen" the collusion[218], there appears to be a clear liability gap compared to older forms of hub-and-spoke cartels, where a single explicit message by the hub can be enough to establish liability, even for spokes that do not reply[219]. TFEU, Article 102 could be invoked[220], but it is not clear what the barriers for this might be, and who would be liable in that case. A similar gap exists in the US as well[221].

As a first step, it seems prudent for antitrust authorities to collect detailed information on the customers of individual pricing algorithm developers, such that they can quickly identify markets which could develop issues around algorithmic hub-and-spokes, and monitor them in more detail. Then potentially problematic dominant pricing algorithms could be investigated, or a limit on the amount of firms who can share a provider of pricing algorithms could be imposed[222]. Presumably, the problematic type of hub-and-spoke collusion unilaterally implemented by the hub would cease to be legal under TFEU, Article 101, once uncovered, as spokes can then "reasonably forsee" future collusive behaviour. Still, they might avoid liability. Potential remedies include additional rules on the liability of third parties for collusive outcomes they facilitate without market participants' knowledge[223], as well as measures to increase firms' ability to foresee collusive efforts by providers of pricing algorithms, such as information campaigns or requirements to regularly self-evaluate their pricing algorithm's decisions for signs of collusion. In addition, sales contracts for pricing algorithms could be prohibited from including profit- or sales volume-based payments, at least for developers with high market shares, in order to remove the incentive for the hub to facilitate collusive pricing.

---

[215] Mehra, Salil K. "Price Discrimination-Driven Algorithmic Collusion: Platforms for Durable Cartels." *Stan. JL Bus. & Fin.* 26 (2021): 171.
[216] In which case it should still expect the algorithm to end up facilitating collusion, as collusion was defined as a strategy to increase joint profits.
[217] Monopolkommission, "Wettbewerb 2018 XXII. Hauptgutachten der Monopolkommission gemäß §44 Abs.1 Satz 1 GWB" (2018) https://monopolkommission.de/images/HG22/HGXXII_Gesamt.pdf
[218] Lorenz Marx, Christian Ritz, Jonas Weller, "Liability for outsourced algorithmic collusion: A practical approximation", *Concurrences N° 2-2019, Art. N° 89925, [www.concurrences.com](www.concurrences.com)* (2019)
[219] Lee, Kenji. "Algorithmic Collusion & Its Implications for Competition Law and Policy." *Available at SSRN 3213296* (2018).
[220] Gautier, Axel, Ashwin Ittoo, and Pieter Van Cleynenbreugel. "AI algorithms, price discrimination and collusion: a technological, economic and legal perspective." *European Journal of Law and Economics* 50.3 (2020): 405-435.
[221] Harrington Jr, Joseph E. "Third Party Pricing Algorithms and the Intensity of Competition." *Available at SSRN 3723997* (2020).
[222] Id.
[223] Monopolkommission, "Wettbewerb 2018 XXII. Hauptgutachten der Monopolkommission gemäß §44 Abs.1 Satz 1 GWB" (2018) https://monopolkommission.de/images/HG22/HGXXII_Gesamt.pdf

### Independent Algorithms scenario

The issues for some instances of the "Independent Algorithms" scenario are already covered in the section on the "Messenger" scenario, as the line between using algorithms as tools for achieving classical tacit collusion, and unknowingly updating a pricing algorithm towards collusion, can be blurry.

Regarding learning agents, we recommend no further immediate actions, as the evidence for their ability to collude real markets is scant. While it is likely that regulatory gaps, for example regarding intent to collude, will emerge once algorithms learn to collude on their own, a cautious approach seems more prudent for various reasons: First, designing good legislation is costly, and policymakers' scarce attention is needed tackling issues with the first two scenarios. Second, some legislation preventing algorithmic collusion in the "Messenger" scenario could have positive knock-on effects on the case involving learning algorithms and potentially reduce the need for additional legislation. Third, our current understanding of if, when, and how algorithms might learn to collude in complex markets is very limited. This is likely going to change over time, especially giving ongoing research on cooperation between AI systems that is being conducted for unrelated reasons[224]. Furthermore, future trends in machine learning and especially learning pricing algorithms are hard to predict, but specific technological details could have an important effect on the effectiveness of different regulations. Thus, regulation is likely to be more effective if implemented, once more is known about the types of learning algorithms that might actually be able to achieve collusion in real markets. Similarly, policymakers might be able to learn from successes and mistakes in regulating current pricing algorithms, if given time to observe the effects of that regulation before designing regulation for learning algorithms. Fourth, premature regulation might hamper innovation and prevent economy-wide efficiency gains from the use of advanced pricing algorithms.

Nevertheless, learning algorithms should not be ignored by antitrust experts. Rather, the use of learning algorithms in a pricing context across the economy should be monitored closely, and policymaker should keep an open eye for stronger and more robust theoretical results on when and how algorithms could learn to collude, as well as empirical demonstrations of successful cooperation between independently trained reinforcement learning agents in complex mixed motive games. Once evidence for the imminent possibility of algorithmic collusion by learning agents begins to mount, the issue will clearly warrant a lot of attention.

## Acknowledgements

We would like to thank Stefan Bechtold for helpful conversations and feedback on this project.

---

[224] Dafoe, Allan, et al. "Cooperative AI: machines must learn to find common ground." *Nature* 593.7857 (2021): 33-36.

# Appendix: Detailed evidence from the literature

The table below shows all papers providing evidence on algorithmic collusion, be it empirical, theoretical, or in the form of informal argumentation, we were able to find in a nonsystematic literature review. We started with keyword searches related to algorithmic collusion in google scholar, and then repeatedly branched out to relevant references cited in the previously identified papers. We only included papers that directly reference AI or algorithmic collusion and ignored papers that merely use RL or other AI algorithms to empirically simulate collusion in a market. This is because we are interested in the effects of AI on collusion, rather than a general exploration of market structures that facilitate collusion for both humans and AI systems (It would be problematic to assume that humans would not be able to implement strategies learnt by an algorithm, without explicit arguments). The papers are generally presented in the order we read them, but papers by the same author are grouped together in order of publication.

Table 1

| Source | Conclusion, Evidence for algorithmic collusion | Type of Evidence | Brief summary and key assumptions | Highlighted features of AI |
|---|---|---|---|---|
| Ezrachi and Stucke 2017[225] "Messengers" Scenario: | Positive, but only for algorithms enabling human collusion. | Informal reasoning, supported by real cases. | AI systems used to monitor and enforce cartel agreements. Example: The Amazon poster case. | Communication is central lin the *Airline Tariff Publishing* example. Speed and efficiency: Automated monitoring and enforcement saves time and effort. |
| Ezrachi and Stucke 2017 "Hub-and-Spoke" Scenario: | Weak positive | Informal reasoning, supported by potential real cases. | Firms either use the same or even a single centralized algorithms to determine prices. (Potential) example: Uber's pricing | Market concentration: A central pricing AI is used. Coordination and credible commitment: Are "outsourced" to the central AI. Symmetry: Copies of the same pricing algorithm might be more likely to "collude" by employing higher, anticompetitive prices. |
| Ezrachi and Stucke 2017 "Predictable Agents" scenario | Quite weak positive: Homogeneous products and few costs to changing suppliers: A collusive Nash-equilibrium might be reached. This might sometimes require automated punishment of defectors. | Informal reasoning | Firms independently make their own pricing more transparent (perhaps to facilitate tacit collusion) and use AI to set prices and predict other's behaviour. | Transparency: Key assumption. Reduces strategic uncertainty and stabilizes tacit collusion. Speed, Efficiency and Learning capabilities: Help to quickly and accurately detect other's price changes and react. This prevents firms from building a reputation as cheaper than others. Symmetry: As in "Hub-and-Spoke", but competitive pressures to use similar algorithms are highlighted. |
| Ezrachi and Stucke 2017 "Digital Eye" scenario | Quite weak positive (scenarios are presented but there is little evaluation of their likelihood) | Informal reasoning | Firms employ advanced self-learning AI systems that learn to tacitly cooperate without human help. | Transparency: Might be found as an optimal strategy by the learning algorithm. Speed and efficiency, Learning capabilities: Enable algorithm to learn and execute the optimal strategy. Symmetry: As in "Predictable Agents". |
| Ezrachi and Stucke 2019[226] | Weak positive: The authors suggest to focus on markets in which tacit collusion already occurs and might be possible without algorithm-to-algorithm communication. This is legitimate and it seems | Informal reasoning | This paper responds to criticism by economists to Ezrachi and Stucke's previous paper. | Implicit and tacit collusion: Algorithmic pricing is expected to make this problem worse. Transparency: Predictable actions of other firms' pricing algorithms, perhaps, as they can be "decoded" could stabilize collusions. Learning capabilities: Complex algorithms with access |

---

[225] Ezrachi, Ariel, and Maurice E. Stucke. "Artificial intelligence & collusion: When computers inhibit competition." *U. Ill. L. Rev.* (2017): 1775.
[226] Ezrachi, Ariel, and Maurice E. Stucke. "Sustainable and unchallenged algorithmic tacit collusion." *Nw. J. Tech. & Intell. Prop.* 17 (2019): 217.

| | | | | |
|---|---|---|---|---|
| | plausible (but far from inevitable, as humans are quite good at coordination) that collusion in these markets will become worse or more likely with pricing set by (learning) algorithms, but the authors' previous writing does not seem to reflect this specific focus, such that the overall argument has the semblance of a motte-and-bailey fallacy. | | | to large amounts of data might be able to distinguish between defection and reactions to market conditions and thus prevent unnecessary retaliation. Speed: Algorithms can retaliate quickly and thus decrease the incentive to deviate from collusion. Commitment: Pricing algorithms could act as a credible deterrent (if their decision are seldomly overruled by humans, which is likely at high speeds). Behavioural consistency: Algorithms are less likely to exhibit human biases, especially in an inconsistent way. They are unlikely to back out of a collusive strategy for fear of detection (unless explicitly programmed to do so). Coordination and Communication: Is (partially) acknowledged as a bottleneck, but the relevance of truly tacit collusion without even machine-to-machine communication is highlighted |
| Miklós-Thal and Tucker 2019[227] (Low uninformed monopoly price) | Mixed: Collusion requires more farsight with better learning capabilities, but affects consumers more negatively, when it happens. | Theoretical: Game theoretic model | Infinitely repeated two-firm price setting game with common discount factor and marginal cost. Sales are split equal for equal prices, but if one firm is cheaper it gets all sales. There are no sales if both firm's prices are above the customers (homogeneous) willingness to pay (wtp). Learning capability is modelled by a common signal providing information about the wtp in the future. The uniformed monopoly price (without observing the signal) is assumed to be small. Collusion is stabilized using grim trigger strategies that permanently drop the price to the marginal cost. | Learning capabilities: Improved demand forecasting leads to higher profits for cartels but also increases temptation to deviate from cartel price. |
| Miklós-Thal and Tucker (High uninformed monopoly price) | Weak negative: Better Learning capabilities make collusion harder. | Theoretical: Game theoretic model | As above, the uniformed monopoly price (without observing the signal) is assumed to be large. | As above. |
| Hansen et al. 2020[228] | Weak positive: Lower noise (better prediction capabilities) leads to higher prices, that eventually reach the monopoly level. | Theoretical/Empirical: RL in model domain, partial validation on real world data. | Two symmetric single product firms, zero marginal cost. Static but noisy, linear (in both prices) demand function. Thus no winner-takes all dynamics. Firms independently learn the demand function by setting prices and observing sales (using the upper confidence bound bandit algorithm). Crucially, firms do not observe other firms' prices such that their demand models are misspecified. | Symmetry: Both firms using the same type of algorithm is crucial for the result, as it relies on both firms mostly setting low prices at the same time and thus not learning that setting low prices can be beneficial to them. Prediction capabilities: Lower noise in the demand could be achieved by conditioning on additional variables. On the other hand, the result seems to depend on bad prediction capabilities, i.e. firms not observing others' prices. |

---

[227] Miklós-Thal, Jeanine, and Catherine Tucker. "Collusion by algorithm: Does better demand prediction facilitate coordination between sellers?." *Management Science* 65.4 (2019): 1552-1561.
[228] Hansen, Karsten, Kanishka Misra, and Mallesh Pai. "Algorithmic collusion: Supra-competitive prices via independent algorithms." (2020).

| | | | | |
|---|---|---|---|---|
| Klein 2019[229] | Quite weak positive: Supra-competitive prices on average. It is not very clear why humans would not tacitly collude under the presented market conditions[230]. | Theoretical/ Empirical: RL in model domain. | Two firms infinitely take turns setting (discrete) prices. Marginal costs are zero. All customers buy at the lowest offered price, and demand is linear in that price.<br><br>Pricing strategies are learnt as a function of the competitor's current price using independent Q-learning, optimizing for cumulative discounted profit. | Speed: The result depends on prices being updated sequentially rather than simultaneously. Interactions of fast AI systems might be better modeled as sequential rather than simultaneous.<br><br>Learning capabilities: Profit for the learnt strategies decreases with finer grained price levels. Better learning algorithms could plausible alleviate this (Q-learning treats each price level as completely independent and does not attempt to generalize). Similarly, the learning rate and the discount rate of the algorithm have relevant effects on performance. |
| Calvano et al. 2020[231] | Weak positive: Supracompetitive profits are achieved on average. It is not very clear why humans would not tacitly collude under the presented market conditions, and it is not clear if the "punishment" found by the authors is actually costly in the short term. Surprisingly, collusion still happens with four firms. | Theoretical/Empirical: RL in model domain. | Infinitely repeated pricing game, simultaneous price setting. Pricing strategies are learnt via independent Q-learning as a function of own and others' past actions, stored in a small memory. Marginal costs are constant. Logit demand, including an outside good[232].<br><br>Discrete prices between a bit below the one-shot competitive equilibrium and slightly above the monopoly price. | Predictive capability: Without looking at the other's prices, or with excessive discounting, the algorithms learn to compete rather than collude. The extent of collusion also depends on various algorithm parameters. In particular finer grained prices again reduce profits.<br><br>Symmetry: The authors find that asymmetry in firms' payoffs (not in the used algorithms, which is only mentioned as an avenue for future research) does reduce the average profit, but less than expected. |
| Han 2021[233] | Unclear: Experience replay (somewhat) reduces the prevalence of collusion for Q-learning, likely as it amplifies problems with non-stationarity. The authors are able to circumvent this with a heuristic modification.<br><br>The authors find no collusion for DQN, both with and without experience replay. However, this result is contradicted by Hettich 2021 (next in table) | Theoretical/Empirical: RL in model domain. | The same model as in Calvano et al. 2020 is used. However, DQN, a more modern version of Q-learning supported by a neural network, and experience replay, where a learning agent reuses past experience instead of only updating on the most recent interaction in order to increase learning efficiency, are used for learning.<br><br>There are some concerns about the paper's quality, in particular technical ones about the DQN results, as the authors claim that most learnt weights remain constant, which likely indicates a flawed implementation of DQN. | Learning capabilities and efficiency: Improvements to the learning efficiency in single agent RL do not necessarily generalize to the multi-agent case and might even reduce performance there. |
| Hettich 2021[234] | Weak positive: As in Calvano 2020, supracompetitive prices are reached. Final profits are slightly lower than in Calvano's experiments, but collusion is learnt faster.<br><br>Profits decline quickly, the more firms there are, but remain slightly supracompetitive, even with ten firms. Again, it is not clear, why humans would not collude to a similar extent in this setting, and whether the "punishment" found by the authors actually qualifies as such. | Theoretical/Empirical: RL in model domain. | Sequential Bertrand model: Each of n firms produces a single product. Quality and marginal costs vary between firms. Demand is given by a multinomial logit model as in Calvano 2020. Price levels are discrete between slightly below the static nash equilibrium and slightly above the monopoly price.<br><br>Firms deploy a version of DQN (modified to optimize average, rather than discounted profit), a modern version of Q-learning that employs a neural network to learn their pricing strategies. Prices are learnt as a function of both firms' prices at the previous time step. | Symmetry: The authors test the slower learning Q-learning against DQN, and find exploitation of the first agent by the second, rather than collusion.<br><br>Learning capabilities: DQN seems to learn to collude more quickly than classical Q-learning (albeit it reaches lower average profits in the end), indicating that the Learning capabilities of algorithms play an important role for algorithmic collusion. The authors also find noticeable effects on joint profits for small adjustments to the algorithms' inputs. |

| Abada and Lambin 2020[235] | Mixed: The authors find "seemingly collusive" outcomes with supracompetitive prices, that are reduced with more market participants. However, the authors find that "punishment" as in Calvano 2020 is also administered for moves to higher prices. They conjecture that this is a sign of suboptimal behaviour due to a lack of exploration, and show that collusion is reduced if exploration is biased towards competition. | Theoretical/Empirical: RL in model domain. | Model based on the electricity market: Firms sell, produce and buy limited amounts of a homogeneous good with price-elastic, linear and time-dependent demand. Unsold goods can be stored without cost, but only up to a firm-specific capacity. In addition, fringe producers only sell the good, at marginal cost. Time is discrete and markets are assumed to always clear. No discounting, but finite time horizon. Demand and production supply are calibrated on data from France.

No direct communication channels exist, and 3 firms attempt to learn strategies (mapping past market prices and the current time to the quantity offered/demanded on the market) maximizing their payoffs using Q-learning with the same parameters. | Learning capabilities: Collusion plausibly gets destabilized by more extensive exploration, which should improve results (at least in single-agent training). This suggests that better learning capabilities could make collusion harder rather than easier in some cases.

Market concentration: The differing levels of collusion for different numbers of market participants are interpreted in terms of market concentration by the authors: the use of shared, central pricing algorithms could reduce the effective amount of sellers in a market and thus facilitate collusion. |
|---|---|---|---|---|
| Ittoo and Petit 2017[236] | Weak negative. Obstacles include understanding other firms' incentives, non-stationarity, scalability and the exploration/exploitation problem, which is harder in the multi-agent case. | Informal/ Theoretical: Analysis of obstacles for RL to reach a collusive equilibrium. Weak empirical support from real world examples. | Different extensions to multi-agent Q-learning could be applied for price setting and others have hypothesized that this could lead to collusive equilibria. | Symmetry: Nash-equilibria are not necessarily reached without coordination on using algorithms designed to converge to them, and equilibrium selection is nontrivial without coordination by the firms.

Learning capabilities: Some previous results showing collusion via Q-learning are not robust to noise. Learning capabilities could reduce this noise. Inferring other firms motivations is highly nontrivial, and it is unclear, how much better firms might get at this. |
| O'connor and Wilson 2021[237] | Mixed evidence: Collusion at the monopoly price is more profitable with AI whenever it is stable. But a (relatively) high likelihood of the predictable demand shock can increase the profitability of deviating. The first effect dominates for sufficiently large discount factors.

Collusion at sub-monopoly prices remains stable with AI, but only negatively affects consumer surplus for (roughly) the parameters for which collusion at monopoly price was enabled by AI. | Theoretical: Game theoretic model | Two firms in infinitely repeated simultaneous pricing game. Common discount factor and static linear demand function subject to two random negative shocks that together reduce demand to zero. Homogeneous goods, zero marginal costs. Customers buy at the lowest offered price.

Neither the other firm's price nor the price shocks are observed directly, such that cheating cannot be identified. Whenever a firm receives no profit it is assumed to punish the other for the minimum amount of rounds that stabilizes collusion.

"AI" makes one the demand shock perfectly predictable. This way, cheating can sometimes be identified, in which case a grim trigger strategy is used. | Learning capabilities: Key assumption. AI sometimes contributes to collusion via better prediction of market shocks. The better learning capabilities allow for sometimes distinguishing cheating from double demand shocks, as one of the demand shocks can sometimes be ruled out. |

| Gautier et al. 2020[238] | Weak negative. Obstacles include exponential complexity growth for multiple firms, non-stationarity, slow learning speed, and generalization.<br><br>Furthermore, most demonstrations of RL-based collusion assume very simple market structures. Lastly, firms might not use RL, but other algorithms. | Informal/ Theoretical reasoning about RL, supported by (the lack of) real world examples for tacit algorithmic collusion. | The lack of examples for tacit algorithmic collusion in the real world, despite theoretical results, is pointed out. It is argued that the "Amazon poster case" falls under explicit collusion. In addition, difficulties in transferring experimental results on algorithmic collusion to the real world are pointed out. | Learning capabilities: Are deemed insufficient for algorithmic tacit collusion in the near term (using (deep) reinforcement learning).<br><br>Symmetry: Is acknowledged, but it is pointed out that the stochasticity of machine learning and asymmetries in data availability could reduce symmetry in practice.<br><br>Implicit/tacit collusion: The seems to dismiss other forms of algorithmic collusion as less relevant, perhaps because non-tacit collusion still requires agreements between firms.<br><br>Transparency: Is seen as hampering collusion, based on Miklós-Thal and Tucker 2019[239]. |
|---|---|---|---|---|
| Beneke and Mackenrodt 2018[240] | Weak positive: Arguments focus on the impact of AI pricing on collusion, but there is little theory or empirical support. | Informal reasoning | Firms employ AI algorithm to predict and set prices. Collusion might ensue in markets that have previously been competitive. | Behavioural consistency: Given sufficiently stable algorithms, accidental price wars caused by irrational human actions could be prevented.<br><br>Speed: Faster price changes and reactions to them reduce profits from undercutting, and might make price signals harder to detect for humans.<br><br>Learning capabilities: Accurate predictions of other algorithms' reactions to price and demand changes might facilitate coordination without the need for communication, especially after demand shifts, and reduce price stickiness which is used as an indicator for collusion.<br><br>Transparency: The "black-box" nature of neural networks might make detection of collusion harder. On the other hand, firms seem to prefer more legible systems. |

---

[238] Gautier, Axel, Ashwin Ittoo, and Pieter Van Cleynenbreugel. "AI algorithms, price discrimination and collusion: a technological, economic and legal perspective." *European Journal of Law and Economics* 50.3 (2020): 405-435.
[239] Miklós-Thal, Jeanine, and Catherine Tucker. "Collusion by algorithm: Does better demand prediction facilitate coordination between sellers?." *Management Science* 65.4 (2019): 1552-1561.
[240] Beneke, Francisco, and Mark-Oliver Mackenrodt. "Artificial intelligence and collusion." *IIC-International Review of Intellectual Property and Competition Law* 50.1 (2019): 109-134.

| | | | | |
|---|---|---|---|---|
| Beneke and Mackenrodt 2021[241] | Very weak positive: While the article is taking algorithmic collusion seriously, its main goal is to discuss legal remedies, rather than providing strong arguments for the severity of the problem. | Informal reasoning | Remedies for algorithmic collusion are identified with an eye to key differences between human and algorithmic collusion | Implicit/tacit collusion: Detection of algorithmic collusion is highlighted as a key problem.<br><br>Learning capabilities: Both pro-competitive (better predictions of the optimal competitive price, more segmentation) and anti-competitive aspects (better prediction of punishing price responses and noise reduction in demand estimates) are discussed. Capabilities might strongly depend on data availability.<br><br>Transparency: The black-box nature of many machine learning models is highlighted as a barrier to detection of algorithmic collusion. On the other hand, transparency is also seen as possibly stabilizing collusion, as competitors might use it to better anticipate price reactions.<br><br>Speed: Very fast punishment of defectors might make it harder to distinguish between oligopoly pricing patterns and competitive pricing patterns. In addition, they can stabilize collusion by removing price lags.<br><br>Behavioural consistency: Irrational price wars out of "anger" might be prevented by using algorithms. Algorithms might also discount the future more consistently.<br><br>Coordination: "Provided that the necessary data is at hand, artificial neural networks could perform well in predicting the price to which all oligopolists in a market are likely to converge"<br><br>Symmetry: Symmetric companies are highlighted as important for collusion, and even the same pricing algorithm might act differently in different firms, as it takes into account cost structures and distribution. |
| Kühn and Tadelis 2017[242] | Weak negative: A key challenge often ignored or dismissed by others is pointed out. | Informal reasoning, supported via citation of model results. | Most arguments for tacit algorithmic collusion ignore coordination problems. | Communication and Coordination: Lack of arguments that algorithms would be good at coordination, and the difficulty of communication between algorithms without explicit programming for it are pointed out.<br><br>Transparency: Sharing algorithms as a coordination mechanism. But they are valuable assets.<br><br>Implicit/tacit collusion: Is highlighted as the main potential problem, but its realism is dismissed, at least for the short term. |

---

[241] Beneke, Francisco, and Mark-Oliver Mackenrodt. "Remedies for algorithmic tacit collusion." *Journal of Antitrust Enforcement* 9.1 (2021): 152-176.
[242] Kühn, Kai-Uwe, and Steve Tadelis. "Algorithmic collusion." *Presentation prepared for CRESSE, URL https://www.ebos.com.cy/cresse2013/uploadfiles/2017_sps5_pr2.pdf f* (2017).

| | | | | |
|---|---|---|---|---|
| Salcedo 2016[243] | Weak positive: When customers arrive a lot faster than firms can decode others' algorithms, or vice versa, collusion necessarily(!) ensues. Either, algorithms are used as "offers to collude", or it is hard to profit from deviations before being punished in the second case.<br><br>However, full transparency seems unrealistic, given competitive incentives to keep superior pricing algorithms private and the black-box nature of neural networks. | Theoretical: Game theoretic model | Two firm price competition with continuous time. Consumers arrive at random times and have random (correlated) willingness to buy each firm's product, and buy at most one product in total. Marginal costs are zero.<br><br>Firms deploy algorithms that set prices based on the market history. Algorithms can be revised, and a firm gains full knowledge of their competitor's algorithm at random times | Commitment: Decoding the other's algorithm is no use if it is immediately changed. If algorithms are only changed seldomly, this is essentially the same as a short-term commitment to a particular strategy. As they are still changed from time to time, commitment is imperfect, which allows convergence.<br><br>Learning capabilities: The result requires that algorithms can implement non-trivial strategies that react to market conditions.<br><br>Symmetry: Asynchronous moves seem crucial to the result.<br><br>Transparency: The result assumes that a firm is able to fully predict the other firm's algorithms counterfactual behaviour for all situations. The authors also claim (but do not seem to prove) that there is incentive to maintain transparency.<br><br>Communication and Coordination: As also pointed out by Kühn and Tadelis, full transparency essentially turns the algorithms into a communication device.<br><br>Implicit/tacit collusion: Is highlighted as a consequence of the result. |
| Calvano et al. 2019[244] | Mixed/very weak positive: The authors highlight many uncertainties but seem to ultimately belief in the plausibility of collusion enabled by algorithms. The authors seem to underappreciate that coordination in mixed-motive games is difficult for RL compared to strong play in zero-sum games such as Chess. | Informal reasoning | Collusion might be enabled by price setting algorithms that are implicitly designed to collude and punish defectors, or by self-learning algorithms that eventually learn to collude. | Speed: Algorithms might react to price changes more quickly than humans, and this is often assumed to facilitate collusion via faster punishment.<br><br>Learning capabilities: The stabilizing property of speed has been called into question when data is noisy[245]. Algorithms might generally be better able to deal with noise. Progress in ML might enable algorithms to effectively collude in complex environments and avoid human biases.<br><br>Communication and Coordination: "Thanks to their proclivity to explore, ML algorithms might solve both coordination problems quite effectively". Algorithms might learn to communicate.<br><br>Implicit/tacit collusion: Potential consequence of algorithms learning to coordinate. |

---

[243] Salcedo, Bruno. "Pricing Algorithms and Tacit Collusion." *Manuscript, Pennsylvania State University* (2016).
[244] Calvano, Emilio, et al. "Algorithmic pricing what implications for competition policy?." *Review of industrial organization* 55.1 (2019): 155-171.
[245] Sannikov, Yuliy, and Andrzej Skrzypacz. "Impossibility of collusion under imperfect monitoring with flexible production." *American Economic Review* 97.5 (2007): 1794-1823.

| | | | | |
|---|---|---|---|---|
| Deng 2018[246] | Mixed/weak negative: The authors find "promising" theoretical and experimental results, but many obstacles to real world deployment. | Informal reasoning | Arguments and theoretical/experimental evidence for and against the plausibility of algorithmic collusion are reviewed. | Symmetry: Symmetry in payoffs Is highlighted as important for collusion. The authors argue that while symmetry in algorithms could facilitate collusion, incentives to customize might be stronger, especially in asymmetrics markets.<br><br>Speed, efficiency and Learning capabilities: "A 'collusive' algorithm is arguably irrelevant to the antitrust community if it takes an unrealistically long time to learn to collude.". The high complexity of real world price setting and the ensuing slow learning is a likely bottleneck for current algorithms. Again, fast interactions in a noisy environment could make collusion difficult[247]. "While a human can certainly check on competitors' prices periodically, a simple web-scraping algorithm can do the same much more efficiently and at a much higher frequency."<br><br>Tacit and implicit collusion: "to design a collusive algorithm, certain "collusive" design features most likely need to be explicitly incorporated into the algorithm. So just like email leaves a trail of evidence when cartels use it to coordinate, a similar trail of evidence is likely present when collusive algorithms are being designed" |
| Assad et al. 2020[248] | Weak positive: Pricing algorithms increase the margins of stations with competitors by roughly 9%, but not for local monopolists. In doupolistic markets, prices increase by roughly 28% if both competitors adopt pricing algorithms, but are unchanged if only one of them does.The gradual onset of the price increases suggests that the algorithms "learn how not to compete", rather than "fail to learn to compete effectively".<br><br>Effects are only significant in the instrumental variable regression. This is problematic as the brand might affect profits via other means. However, similar results are obtained using local internet speed as an instrument. | Empirical: Real world effects of pricing algorithms are analyzed. | Algorithmic pricing became widely available in the german gasoline retail market in 2017. The authors use structural breaks in markers associated with algorithmic pricing (frequency and size of price changes, delay in reactions to rivals' price change), to estimate station-level and brand headquarter-level adoption of algorithmic pricing. Then, brand-level adoption is used as an instrument to estimate the effect of station-level adoption on price shifts. | Symmetry: The extent to which the gas stations are using the same or different algorithms is unclear.<br><br>Institutions: "Most algorithmic pricing software are "cloud" based and require constant downloading and uploading of information.". This could make a "Hub-and-Spoke" scenario more likely.<br><br>Speed: "High-frequency interaction between competitors can make collusion easier to sustain due to increased ease of monitoring and quicker detection and punishment of deviations"<br><br>Learning capabilities: Likely play an important role, as indicated by the delayed onset of price increases after the adoption of pricing algorithms. |
| Zhou et al. 2018[249] | Very weak positive: Humans learn (in human-human interactions) to collude in the same model, but more slowly[250]. In addition, the algorithm seems to mostly rely on an explicit theoretical collusion strategy rather than self-learning. Lastly, grammatical errors reduce readability, and supplementary material is missing. | Theoretical/Empirical: Game theoretical modelling, results validated in human vs algorithm experiment. | Two firms selling the same product. Firms simultaneously decide on their production quantity in each round of an iterated game. The price is a decreasing function of the quantities produced by both firms. Marginal costs are constant.<br><br>The authors design a parameterized algorithm, based on a theoretical collusion strategy, that learns to extort a human player into collusion. | Implicit and tacit collusion: Is highlighted as problematic and plausible because of the result.<br><br>Behavioural consistency: It seems likely that the consistent punishment strategy used by the algorithm increases the speed at which the human decides to collude. |

| Schrepel 2019[251] | Positive regarding the potential impacts of the blockchain and smart contracts on the stability of collusive agreements. However, some important caveats, such as the weak link between the real world and the blockchain for verified information (how does the accurate information get to the blockchain in the first place?) seem to be ignored. | Informal reasoning | This paper investigates the potential impact of the blockchain and smart contracts on competition. | Commitment: Binding commitments and trust can play a fundamental role in stabilizing collusive agreements. The blockchain can be used to improve trust via the provision of easily verifiable information, or to create credible commitments to certain pricing strategies via smart contracts.<br><br>Transparency: As mentioned above, the blockchain could be used to provide accurate information and thus provide market transparency. In addition, knowing the strategies others are committed to via smart contracts can also be seen as a form of transparency. On the other hand, private blockchains can make it difficult for authorities to identify participants in a collusive agreement.<br><br>Implicit and tacit collusion: While not discussed directly, explicit agreements to collude that are hard to uncover and prove might cause similar problems as tacit collusion. In particular, blockchain applications might make it easy to quickly exit cartels that are in danger of identification. |
|---|---|---|---|---|
| Schrepel 2020[252] | Mixed: Regarding non-blockchain-based algorithmic collusion, the authors refer to the lack of empirical evidence and quantification, and claim that algorithmic collusion does not create any fundamental issues, but their arguments lack detail. Details are also lacking regarding blockchain, but its ability to facilitate commitment seems very important. | Informal reasoning | This article argues that the saliency of algorithmic collusion is inflated due to publication bias, and that blockchain-based collusion (which does not seem to usually be discussed in papers on algorithmic collusion) is more problematic. | Commitment: Is highlighted as key difference between blockchain-based and other types of algorithmic collusion: "The selling of a product at a price different than the one agreed upon by colluders could be recorded automatically into the blockchain by way of smart contracts, and these smart contracts could punish the behavior [...] As smart contracts are immutable, no colluders can change the set governance. |
| Bernhardt and Dewenter 2020[253] | Weak negative: Communication between algorithms and transparency are highlighted as key near-term bottlenecks for full-blown algorithmic collusion. Algorithms as tools for explicit collusion (as in Ezrachi and Stucke's "messenger" or "Hub-and-Spoke" scenario, are deemed more realistic. | Informal reasoning | This paper discusses both algorithmic collusion and "collusion by code", in which computer code is used to enforce collusion. | Symmetry: The use of similar algorithms is highlighted as facilitating collusion.<br><br>Transparency: Both market transparency and algorithms being able to decrypt other algorithms are highlighted as potentially facilitating collusion. However the former might also help consumers.<br><br>Learning capabilities and efficiency: The need for both advanced algorithms and large amounts of data (that has to be processed efficiently) to actually obtain high degrees of transparency is highlighted. However, better learning algorithms might increase the consistency of punishment, as algorithms might make fewer error than humans.<br><br>Speed: This allows firms to punish deviations more quickly.<br><br>Coordination and communication: Pricing algorithms might make coordination easier, especially if they are able to communicate with each other.<br><br>Implicit and tacit collusion: Is discussed as a severe but unlikely problem. |

---

[251] Schrepel, Thibault. "Collusion by blockchain and smart contracts." *Harvard Journal of Law and Technology* 33 (2019)
[252] Schrepel, Thibault. "The Fundamental Unimportance of Algorithmic Collusion for Antitrust Law." *Harvard Journal of Law and Technology* (2020).
[253] Bernhardt, Lea, and Ralf Dewenter. "Collusion by code or algorithmic collusion? When pricing algorithms take over." *European Competition Journal* 16.2-3 (2020): 312-342.

| | | | | |
|---|---|---|---|---|
| Brown and MacKay 2021[254] | Positive: In the price setting model, prices are higher than with simultaneous price setting. In the algorithm model, the same is true as long as both algorithms respond to price increases or decreases in the same kind. In addition, there exists a linear(!) algorithm that supports fully collusive prices. | Theoretical/Empirical: Game theoretic model, built on empirical pricing data. Simulation results. | The authors first analyze hourly price data for allergy drugs in five US online retailers from 2018 and 2019 and find price updates at regular intervals that differ between firm, as well as faster firms quickly reacting to price changes of slower firms and having consistently lower prices.<br><br>Then, a model is built reflecting these facts: Two firms operate in a market with consumers arriving continuously, at a constant (stochastic) rate. Profit is discounted at a constant rate and quasiconcave in prices. Products are substitutes. One firm is able to update their prices each unit of time, and the other more frequently. Lastly, prices are strategic complements (the best response price of a firm is larger for larger prices of the other firm). A similar model is used for simulation in a three-firm oligopoly.<br><br>In another model, pricing algorithms (that determine prices based on others' current prices) are chosen instead of prices and updated at a lower frequency than the prices. This time, both firms update their algorithms at the same time, and the algorithms are both able to update prices at very high frequencies. Coordination problems between algorithms are assumed to be solved by an adversarial market coordinator who selects the equilibrium with the lowest profit. Lastly, algorithms are restricted to continuous functions of others' prices to avoid "obvious" punishment that is easy to detect. | Commitment: Credible commitment (via an algorithm or price that is fixed in the short term) is essential to supracompetitive outcomes in both models.<br><br>Symmetry: Asymmetric pricing frequencies are key to the price increase in the first model, as the essentially allow for short-term commitments by the slower firm. In particular, a firm might have incentive to keep its pricing process slower to maintain supracompetitive profits, even if it earns less than the other firm, due to its slowness.<br><br>Efficiency, Learning capabilities and behavioural consistency: Combined with ample price data in online markets, algorithms can be used to make faster pricing decisions at scale and stick to a consistent strategy better than error-prone humans. Observing and being able to take into account the other firm's price is crucial to supracompetitive pricing in the second model. Even a simple linear algorithm can maintain collusion. However, more complicated algorithms with memory might face similar difficulties to collusion as humans do.<br><br>Speed: Higher speed compared to rivals or the frequency of algorithm updates exacerbates the respective commitment effect.<br><br>: |
| Ballard and Naik 2017[255] | Very weak positive: The authors focus on the potential for algorithmic collusion, but their arguments lack detail. | Informal reasoning | The authors discuss current examples related to algorithmic collusion and speculate about future problems related to AI and collusion. | Implicit and tacit collusion: Difficulties with detecting algorithmic collusion compared to human collusion are highlighted.<br><br>Speed, and behavioural consistency: Are highlighted as key advantages of algorithms.<br><br>Learning capabilities: Fast progress in AI is cited to support the relevance of algorithmic collusion, as it might result in more efficient strategies being learnt faster.<br><br>Communication and coordination: Are stipulated as a unique capability of AI. |
| Petit 2017[256] | Weak negative: Several limitations to the literature on algorithmic collusion are highlighted: A lack of focus on how algorithmic pricing can hinder collusion, how (company) customers can use technology to defend themselves against cartels, as well as empirical evidence and modelling results with realistic assumptions | Informal reasoning | Research agenda on antitrust and AI | Symmetry: Is often assumed to prove algorithmic collusion, but unrealistic because of competitive pressures to develop better algorithms and ongoing algorithm research.<br><br>Tacit and implicit collusion: Is mentioned as a key argument in most research on algorithmic collusion.<br><br>Learning capabilities: Non-price competition on privacy and behavioural discrimination by pricing algorithms could increase noise (if not all data is publically available) and make it more costly to react to perceived deviations.<br><br>Behavioural consistency: It is acknowledged that a lack of fear of detection and penalties for collusion could make algorithms more likely to collude, but it is argued that firms won't want to use such algorithms, as they do want to prevent detection and penalties. |

---

| | | | | |
|---|---|---|---|---|
| Azutti et al. 2021[257] | Very weak positive: The authors highlight several obstacles to the use of machine learning in trading and point out the lack of empirical evidence for algorithmic collusion, but remain positive about the future feasibility. They also argue for the insufficiency of current legislation to deal with tacit algorithmic collusion. | Informal reasoning, | This paper analyses the effects of machine learning on (financial) market integrity, specifically market manipulation and tacit collusion. It then analyzes, how the black-box nature of many modern machine learning systems could make it hard to prevent tacit collusion under the current antitrust framework. | Tacit and implicit collusion: Is highlighted as a key problem with algorithmic trading, as firms might be able to avoid liability for collusive behaviour.<br><br>Transparency: The black box nature of modern machine learning systems is highlighted as a key problem for dealing with tacit algorithmic collusion. On the other hand, market transparency in digital marketplaces (in which AIs are more likely to operate first) is seen as facilitating collusion. The same goes for transparency between algorithms, which allows for predicting others' strategies.<br><br>Efficiency and Learning capabilities: AI's ability to process large amounts of data is highlighted as a key advantage over humans. Further improvements in Learning capabilities are seen as a key requirement for real algorithmic collusion.<br><br>Behavioural consistency: The authors (perhaps mistakenly) argue that machine learning methods are "simply constrained by the empirical nature of data", while humans rely on intuitions and gut-feelings.<br><br>Speed: Trading frequency is an important factor for collusion as it allows for swift punishment of defectors and can be increased using algorithms.<br><br>Coordination and Communication: Are key capabilities for algorithmic collusion, but algorithmic capabilities in these domains might still be too limited. |
| Schwalbe 2018[258] | Weak negative: Only few modelling studies are seen as demonstrating algorithmic collusion, and most of them require conditions in which humans have been demonstrated to collude more efficiently. There does not seem to be evidence for complex neural network-based algorithms to learn to collude in complex mixed-motive games. | Informal reasoning supported by citing results from both game theoretic and real-world experiments. | While legal scholars often assume that tacit algorithmic collusion is easily achieved, computer scientists and economists hold against this that tacit collusion is not always achievable, and both hard and slow to achieve, when it is. This paper (seemingly comprehensively) reviews both literatures and highlights communication as a key bottleneck for achieving algorithmic collusion. | Communication and coordination: Communication is seen as vital, but the ability of learning algorithms to communicate (which is particularly important in oligopolies with more than two firms) is very limited at the moment, but might improve in the future. The difficulty of coordination also strongly depends on the market structure and other details.<br><br>Learning capabilities: AIs' ability to handle complex and dynamic markets is highlighted as a key current bottleneck for tacit algorithmic collusion. On the other hand, complex algorithms do not seem to fare well in game theoretic experiments. Better data processing could increase market transparency and increase the ease of detecting deviations from collusive prices.<br><br>Transparency: Algorithmic transparency can be used to facilitate communication.<br><br>Speed: Swifter punishment by fast algorithms could decrease the incentive to deviate from collusive prices.<br><br>Market concentration: Economies of scale around pricing algorithms might increase market concentration, but these algorithms are often offered by independent firms.<br><br>Symmetry: The focus of most experimental studies on symmetric (both algorithmically and in terms of market position) scenarios, which might not be given in the real world is highlighted.<br><br>Tacit and implicit collusion: Learning, rather than static algorithms are important for tacit algorithmic collusion. Collusion might strongly depend on idiosyncratic details of the employed learning algorithms. |
| Mehra 2015[259] | Weak positive: The paper reviews evidence for the wide adoption of pricing algorithms and provides some arguments for why | Informal reasoning | This article analyses the consequences of an increased adoption of algorithmic pricing, with a particular focus on consumer welfare and antitrust law. | Speed: Faster punishment of deviations from collusive prices can stabilize cartels.<br><br>Learning capabilities: Can reduce the negative effect of noisy market estimates on collusion, as accidental |
| | algorithmic pricing could facilitate collusion. Unsurprisingly, given the early publishing date of the article, many key obstacles to algorithmic collusion that were later pointed out by others are ignored. | | | punishment of others' reactions to unanticipated/unobserved demand shifts can be avoided. On the other hand, real defection becomes easier to detect and punish with more data availability.<br><br>Behavioural consistency: Using pricing algorithms avoids temporally inconsistent hyperbolic discounting. In addition, principal agent problems with executives who benefit from short term price decreases but might be unaffected by the punishment can be avoided.<br><br>Implicit and tacit collusion: Anthropomorphic concepts like "intent" play an important role in antitrust law, but cannot be applied to pricing algorithms.<br><br>Communication, coordination and commitment: The authors argue that communication might lose its importance for trust building, when algorithms are able to punish defectors immediately for the smallest infringements in order to demonstrate their commitment to a punishment-based strategy.<br><br>Market concentration: Can be a consequence of centralized pricing algorithms, such as Uber's. |
| Mehra 2021[260] | Positive, but only for hub-and-spoke type collusion. | Informal reasoning/ Game theoretical modelling | This paper explores synergies between price discrimination and algorithmic collusion in a hub-and-spoke fashion. | Learning capabilities: Big data and modern algorithms allow for more precise estimates of customers' willingness to pay, and thus price discrimination.<br><br>Market concentration: Hubs (perhaps Uber) use their access to data and advanced algorithms to enable spokes (Uber drivers) to price discriminate more efficiently than on their own. This way, more and more spokes concentrate at the same hub, which can use this market power to set the price for the spokes' services above the competitive level. Crucially, the increase in gains via price discrimination can enable collusion in cases where firms would otherwise be incentivized to defect from using the hub for pricing.<br><br>Coordination: Is done automatically via the hub, such that large numbers of participants might be able to collude. |
| Competition and Markets Authority 2018[261] | Mixed: The authors do find evidence that pricing algorithms are widely used to set prices online, and state that pricing algorithms could strengthen existing collusive agreements. However, the relevance of modelling results on tacit collusion is drawn into question, in part because they usually treat the type of deployed algorithm as exogenous, rather than allowing firms to deviate by using more sophisticated algorithms.<br><br>The authors find algorithmic hub-and-spoke collusion, and collusion in "marginal markets", where collusion is already almost possible to be more worrying compared to other scenarios, for which the evidence is lacking.<br><br>Lastly, they find a tension between algorithmic collusion and personalized pricing, | Informal reasoning | Report by the United Kingdom's Competition and Markets Authority on pricing algorithms. | Learning capabilities and efficiency: Algorithms can efficiently process large amounts of data and thus better identify deviations from collusive agreements. They can also cheaply adjust large amounts of prices. Better learning capabilities could be used to better differentiate between "legitimate" responses to demand shifts and deviations from collusion, which helps avoiding price wars. In addition, they could help with identifying the price to maximize joint profits. Very advanced algorithms might even make it possible to coordinate on (personalized) prices for many goods.<br><br>Speed: Fast punishment of deviations from collusive agreements enabled by pricing algorithms can stabilize collusion<br><br>Transparency: Predictable behaviour by (simple) pricing algorithms can stabilize collusion. However, most machine learning algorithms are better characterized as black-boxes rather than transparent.<br><br>Coordination and communication: The authors link the process of decoding other firms' algorithms to explicit communication of questionable legal status. They also highlight the need of communication to maintain collusion in markets with personalized pricing. Lastly, they do highlight the potential for algorithms to use price signals as a tools for coordination.<br><br>Symmetry: Symmetry between competitors' algorithms |

---

| | | | | |
|---|---|---|---|---|
| | | | | is claimed to increase the likelihood of collusion, in part because it allows firms to better predict each others' responses. However, it might not be trivial for firms to know they are using the same algorithm without resorting to explicit communication that could be used as evidence for collusion

Behavioural consistency: Unlike human employees, algorithms might not destabilize agreements by undercutting prices for short-term profits linked to promotions or boni.

Market concentration: A common intermediary providing its pricing algorithm to many firms in the same market could act as a hub facilitating collusion. Algorithms might increase barriers to entry a market by quickly identifying and pricing out new entrants, perhaps only targeting likely customers of the entrant. |
| Ohlhausen 2017[262] | Weak negative: Algorithms that send price signals are framed as yet another form of illicit but hard to detect communication, similar to the move from paper to email. Similarly, algorithmic hub-and-spokes are seen as clearly illegal under current antitrust law. | Informal reasoning | Speech by the (former) acting chairman of the US Federal Trade Commision on | Speed: Is highlighted as a key advantage of algorithms

Prediction capabilities: Is highlighted as a key advantage of algorithms

Communication: Algorithms might be used to send and interpret price signals

Market centralization: The possibility of Hub-and-Spoke scenarios via algorithmic pricing is highlighted, but these are framed as not a novel problem, that can be dealt with. |
| Harrington 2018[263] | Mixed: The lack of evidence from real markets is pointed out, but the possibility is held up, due to modelling results and the general difficulty of predicting the future of IT. | Informal reasoning | This paper reviews legal challenges to dealing with collusion by learning algorithms and proposes several remedies. | Implicit and tacit collusion: Is seen as a key challenge that might become problematic with advanced pricing algorithms.

Transparency: Employee's inability to predict a learning algorithm's behaviour is pointed out as an important challenge to keeping firms liable. On the other hand, a lack of market transparency in personalized pricing is seen as a potential roadblock to algorithmic collusion (If personalized prices are not observed, deviations from collusive prices are harder to punish).

Prediction capabilities: A strong ability to perceive he market environment is seen as an important factor to enable collusion by learning algorithms. On the other hand, the non-stationarity of multi-agent interactions, as well as the noise introduced by exploration is highlighted as a key problem for efficient learning. |
| Harrington 2020[264] | Weak positive: Prices rise, even when only one firm adopts the algorithm, as the possibility that others could have adopted the algorithm incentivizes it to (implicitly) value the other firm's profits. Both the higher the probability of adoption, and the higher the number of actually adopting firms, the stronger this effect. | Theoretical: Game theoretic modelling | Two firms with differentiated products. Marginal costs are constant and a firms' profits are strictly concave in its price. Demand is linear, decreasing in a firm's own price and increasing in the other firm's price, but also affected by another outside variable, about whose distribution both firms have a common belief.

Third-party pricing algorithms are assumed to observe the outside variable such that they are better able to predict demand. Firms' decisions to adopt the algorithm are assumed to be independent and the algorithm and its outputs are assumed to not directly depend on how many firms have adopted it (to prevent obvious | Tacit and implicit collusion: The possibility of tacit collusion, in form of am implicit hub-and-spoke cartel is the main issue found in this paper.

Communication and Coordination: The algorithms unawareness of its adoption by other firms is chosen to avoid antitrust authorities interpreting such awareness as illicit communication.

Market concentration: Is crucial here. Even the possibility of high market concentration causes price increases in the model, as the algorithm acts more like a hub-and-spoke, just in case it might have high market concentration.

Prediction capabilities: Stronger prediction capabilities of third parties, due to specialization and access to |

---

[262] Ohlhausen, Maureen K. "Should We Fear The Things That Go Beep In the Night? Some Initial Thoughts on the Intersection of Antitrust Law and Algorithmic Pricing." Federal Trade Commission (2017) https://www.ftc.gov/system/files/documents/public_statements/1220893/ohlhausen_-_concurrences_5-23-17.pdf

[263] Harrington, Joseph E. "Developing competition law for collusion by autonomous artificial agents." *Journal of Competition Law & Economics* 14.3 (2018): 331-363.

[264] Harrington Jr, Joseph E. "Third Party Pricing Algorithms and the Intensity of Competition." *Available at SSRN 3723997* (2020).

| | | | collusion). On the other hand, it is allowed to depend on the probability of adoption, and non-adopting firms can change their prices in response to others' adoption. The algorithm is built to maximize a firm's profit given these assumptions. | more data, facilitate market concentration. |
|---|---|---|---|---|
| Lee 2018[265] | Very weak positive: The paper cites evidence for the prevalence of algorithmic pricing and interprets the existing evidence on collusion facilitated by non-learning algorithms as suggestive. While the authors do acknowledge the lack of strong evidence for collusion by learning algorithms, they do deem it a "real", rather than a "theoretical" risk. | Informal reasoning | This paper reviews and analyses the literature from (singaporian) law, economics and computer science relevant to algorithmic collusion. | Implicit and tacit collusion: Is highlighted as a problem, especially for learning agents.<br><br>Learning capabilities, and efficiency: Market transparency is increased by business employing algorithms to conform to competitive pressures. The ability of algorithms to process large amounts of data efficiently is important for this. But market transparency allows firms to more easily detect and punish deviations from collusive agreements.<br><br>Speed: Algorithms allow for the swifter punishment of deviations from collusive prices.<br><br>Behavioural consistency: The lack of human biases in algorithms might reduce firms' propensity to "cheat" and set noncollusive prices.<br><br>Market concentration: Algorithms might use limit pricing to discourage market entry or eliminate existing competitors. |
| Siciliani 2019[266] | Very weak positive: The authors note that in their model, no firm has an incentive to ever lower prices below the monopoly price and claim that monopoly pricing is the only Nash-equilibrium even in the (not iterated) one-stage game. However, they do not prove this formally, and it remains unclear why competitive pricing should not be a (pareto-dominated) nash-equilibrium as well. Also the authors note that once a single firm does not use the pricing algorithm, collusion becomes more complicated. | Informal reasoning (very simple) game theoretical modelling | Bertrand competition model: Products are close substitutes, search and switching costs for customers are limited, firms do not face capacity constraints. Pricing algorithms are assumed to immediately match others' price decreases. | Speed: Instantaneous price matching eliminates short-term gains from undercutting rivals.<br><br>Market concentration: The potential role of e-marketplace operators in facilitating hub-and-spoke type collusion (not by using a central pricing algorithm, but by omitting to limit the frequency of price updates) is discussed, but seen as less problematic as long as there is between-platform competition (buyers are likely to choose the platform that prevents monopoly pricing).<br><br>Implicit and tacit collusion: While not discussed in detail, pointing out the potential for legal tacit collusion stabilized by pricing algorithms appears to be part of the author's goals. |
| Marx et al. 2019[267] | Weak positive (but only for algorithms being used as tools for explicit collusion): Multiple (alleged) examples are presented for the first two scenarios | Informal reasoning | Liability is discussed for three scenarios Ezrachi and Stucke's "messenger", "hub-and-spoke", as well as a new "invisible hand" scenario similar to "hub-and-spoke", but in which the "hub" (developer of a widely used pricing algorithm) facilitates collusion without the "spokes'" knowledge. | - |
| O'Kane and Kokkoris 2020[268] | Weak positive: Example cases in which algorithms were used for explicit collusion are provided. The authors point out that the lack of cases for the latter two scenarios might indicate difficulties with bringing such cases to court. | Informal reasoning | Three scenarios are discussed: Algorithms as a vehicle to implement an existing agreement, formation of an alliance via a third party algorithm, and alignment via the "plain interaction and adaption" of pricing algorithms. | Learning capabilities: The accurate identification of price deviations is important for using algorithms as a vehicle to implement collusive agreements. Strong Learning capabilities would also be necessary for algorithms to learn to collude.<br><br>Market concentration: Plays an important role for the scenario involving a shared third-party algorithm.<br><br>Implicit and tacit collusion: Is mentioned as a problem, in particular for the last scenario. |

---

[265] Lee, Kenji. "Algorithmic Collusion & Its Implications for Competition Law and Policy." *Available at SSRN 3213296* (2018).
[266] Siciliani, Paolo. "Tackling Algorithmic-Facilitated Tacit Collusion in a Proportionate Way." *Journal of European Competition Law & Practice* 10.1 (2019): 31-35.
[267] Lorenz Marx, Christian Ritz, Jonas Weller, "Liability for outsourced algorithmic collusion: A practical approximation", *Concurrences N° 2-2019, Art. N° 89925, www.concurrences.com* (2019)
[268] O'Kane, Claudia and Kokkoris, Ioannis. "A Few Reflections on the Recent Caselaw on Algorithmic Collusion." *Competition Policy International, Antitrust Chronicle* (2020).

| | | | | |
|---|---|---|---|---|
| Blockx 2017[269] | Mixed: While some arguments for the possibility of algorithmic collusion are briefly laid out, they lack detail. On the other hand, the paper is optimistic about EU antitrust law's capability to handle algorithmic collusion. | Informal reasoning | This paper discusses algorithmic collusion and tools from EU antitrust law that can be used to prevent it. | Learning capabilities: Differences between consumer and algorithm-aided sellers' ability to monitor the market could lead to anti-competitive outcomes.<br><br>Market concentration: Is mentioned as a potential anticompetitive factor in digital markets.<br><br>Implicit and tacit collusion: Is seen as a key problem.<br><br>Behavioural consistency: Human irrationality is seen as helping the detection of anticompetitive practices, and algorithms presumably lack this irrationality. |
| Chen et al. 2016[270] | Positive regarding the prevalence of algorithmic pricing: Over 500 sellers using algorithmic pricing are identified, and the authors find evidence that these sellers might have larger sales volumes. Interestingly, algorithmic sellers seem to sell a smaller range of products, compared to other sellers.<br><br>The authors do not find any evidence of market distortions by algorithmic sellers. | Empirical: The prevalence and effect of algorithmic pricing on the Amazon marketplace is analyzed. | The authors use web scraping to collect data for 1641 popular products on Amazon and the top 20 sellers for each of these products, every 25 minutes for four months. Prices are compared to three potential target prices: The lowest price, Amazon's own price and the second lowest price. Sellers that a) frequently change their prices and b) exhibit strong similarity to one of the target prices, are presumed to use pricing algorithms. | Efficiency and speed: Algorithmic sellers are found to change prices at high frequency for multiple produces, which could not be done by humans<br><br>Market concentration: The authors find indications of a competitive advantage for sellers using algorithmic pricing, which could force more and more sellers to use algorithms, over time. This seems to be exacerbated by the workings of the Amazon marketplace, which can exhibit winner-takes-all dynamics. |
| Monopolkommission 2018[271] | Mixed: The report points out that the effects of pricing algorithms are hard to forsee and cites the strong strong dependence of market structures on the effect of pricing algorithms, but points out that the internet economy might be particularly suitable for (tacit) algorithmic collusion. | Informal reasoning | Report on algorithmic collusion by the german monopoly commission, an independent committee on antitrust law that takes an advisory role for the german government. | Speed: Pricing algorithms can adapt prices quickly and thus facilitate quicker punishment of deviations from collusive prices.<br><br>Learning capabilities and efficiency: Better information about competitors' behaviour trough learning algorithms and big data can help to stabilize collusion.<br><br>Tacit and implicit collusion: Algorithms might facilitate collusive outcomes without explicit agreements. Learning algorithms can temporally disconnect the human decision leading to collusion and the collusive behaviour.<br><br>Transparency: "Obfuscation" of algorithm code to make understanding the decision process harder is highlighted as a problem to proving algorithmic collusion.<br><br>Communication and coordination: Algorithms could be used to automatically send price signals.<br><br>Market concentration: Hub-and-spoke type collusion can be facilitated by central price platforms using a common algorithm. |
| OECD Secretariat 2016[272] | Weak positive: The authors seem to mostly draw on Ezrachi and Stucke 2017's arguments and examples in the papers' section about antitrust. However, they additionally provide strong arguments for increased market concentration through big data. | Informal reasoning | This paper looks at the effects of big data in a competition context. | Market concentration: Various arguments for a strong effect of algorithm use in big data on market concentration are brought forward: Network effects that cause data availability and service quality to enter a positive feedback loop, as well as cost structures with high initial and low marginal costs.<br><br>Learning capabilities and speed: Are argued to further facilitate market concentration by allowing for better and faster "nowcasting", and using the improved information availability to prevent new market entries.<br><br>Implicit and tacit collusion: Is highlighted as a problem, |

---

| | | | | | |
|---|---|---|---|---|---|
| | | | | | and the low number of cases against digital cartels is interpreted strengthening the incentive to form them, rather than as evidence against their existence. |
| Bundeskartellamt and Autorité de la concurrence 2019[273] | Weak negative: The paper concludes that the likelihood of tacit algorithmic collusion in real markets remains an open question. In particular, the relevance of theoretical results on algorithmic collusion are called into question, based on the amount of necessary assumptions.<br><br>The authors also stipulate that current EU legislation might be adequate for dealing with algorithmic collusion, and highlight that the nature of competition issues due to algorithmic pricing is not yet known well enough to warrant action. | Informal reasoning | Joint report by the german and french antitrust authorities on the effect of algorithms on competition. Three scenarios are discussed: Algorithms used to support existing collusive agreements, collusion via a shared third-party algorithm, and collusion via the parallel use of individual algorithms. | Coordination and communication: The potential for communication between self-learning algorithms is seen as uncertain, but price-signalling is highlighted as a potential means of communication that could facilitate collusion.<br><br>Speed: Quick punishment of price deviations by algorithms might stabilize cartels. Speed might also allow for reducing the cost of price signals, by only changing prices for short times that consumers cannot exploit.<br><br>Learning capabilities and efficiency: Efficiently and accurately monitoring others' prices and automating price adaptions through big data and algorithms could facilitate collusion. The cost of monitoring compared to humans might be an important factor. Better exploration strategies might increase the likelihood that algorithms find collusive strategies. In addition, better learning capabilities might improve transparency. Lastly, potential losses due to algorithms learning slowly could be an obstacle for algorithmic collusion via self-learning algorithms, especially if market conditions are in constant flux. While complex algorithms might have better collusive capabilities at convergence, they might also learn more slowly.<br><br>Transparency: Collusion might be easier to initialize if the behaviour of others' can be understood more easily through the use of advanced algorithms. On the other hand, algorithms might be able to mask collusion by deliberately introducing indicators of competition, such as price heterogeneity and price instability, from time to time.<br><br>Implicit and tacit collusion: Might or might not arise due to pricing algorithms. If it does, it might not be illegal under current european antitrust law.<br><br>Market concentration: Links between the use of algorithms and firms' market power might create additional barriers to entry. However, the existence of third-party pricing algorithms and data brokers might alleviate this. But then, market concentration for third-party algorithms could facilitate hub-and-spoke collusion.<br><br>Symmetry: Might play a role for algorithmic collusion, but might not be given in the real world (both regarding algorithms and firms' market position). |
| OECD 2017[274] | Weak positive: The paper assesses tacit algorithmic collusion as a serious problem, and provides examples for algorithms being used as tools for collusion. However, it cautions to first develop a better understanding of the issue before taking actions. The lack of empirical evidence for more complicated forms of algorithmic collusion is pointed out, but this is in part attributed to the difficulty of detection.<br><br>The modelling part confirms | Informal reasoning, simple game theoretical modelling | OECD report on algorithmic collusion.<br><br>The game theoretical model assumes that deviations from a collusive price can be detected and punished instantly, and finds that collusion can always be sustained as an equilibrium strategy. | Speed: Is pointed out as a key distinguishing feature of algorithms that can facilitate collusion. Increased speed might also make price wars or punishment strategies harder to observe for humans, and decrease the expected cost of price signals that failed to facilitate coordination.<br><br>Implicit and tacit collusion: Is an important concern throughout the paper. In particular, the potential for managers to collude via self-learning algorithms without the need for explicit communication between the managers is highlighted.<br><br>Learning capabilities and efficiency: Powerful algorithms could be used to efficiently monitor prices and increase market transparency, to punish deviations from collusive agreements, or to learn to optimise joint profits without the need for humans to agree on collusion. |

---

| | | | | |
|---|---|---|---|---|
| | that faster detection of deviations from collusive prices and faster punishment reactions can make collusion possible where it was not before. | | | Coordination and communication: Price signals could be used to coordinate complex cooperative strategies. Algorithms might reduce associated costs, as price signals could be made too short-lived to incur costs if others fail to adjust their prices. The authors seem to assume that self-learning algorithms would be able to coordinate on equilibrium selection: "Even when an infinite number of anti-competitive prices can be sustained, it is also likely that self-learning algorithms can more easily determine the price that maximises joint profits"<br><br>Symmetry amd market concentration: Firms could use copies of the same algorithm programmed to maximize joint profits. Similarly, a third party to which most firms' pricing decisions are outsourced could fulfill a similar role. |
| Gal 2019[275] | Weak positive: The authors highlight the possibility for algorithms to facilitate coordination, even though they admit that this is not inevitable. They acknowledge the low levels of current enforcement regarding algorithmic collusion, but maintain that this might be due to adoption delays in antitrust agencies, rather than a lack of anticompetitive conduct.<br><br>Still, they recommend caution regarding novel regulation and a primary focus on the most obvious cases of algorithmic collusion. | Informal reasoning | This article investigates the possibility of algorithmic collusion and the extent to which both market solutions and current antitrust law can effectively deal with this. | Learning capabilities and efficiency: Advanced algorithms might be able to quickly learn how to coordinate, perhaps even before being deployed to actual markets. Additionally, collusive strategies could be adapted to changing market conditions more quickly with better algorithmic capabilities. Algorithms' ability to process large amounts of data efficiently could also enable them to better predict competitors' strategies and more accurately distinguish deviations based on changing conditions from "defection".<br><br>Transparency, coordination and communication: The authors repeatedly highlight the ability of algorithms to "read the minds" of other algorithms, and how this novel mode of communication could facilitate coordination. In addition, just publicly announcing the use of any pricing algorithm could constitute a market signal providing information about what kind of pricing strategy is employed. On the other hand, algorithms might be able to obfuscate communication between them, making collusion harder to detect.<br><br>Implicit and tacit collusion: Is highlighted as an important legal challenge posed by pricing algorithms.<br><br>Commitment: If algorithms are difficult or costly to change, they essentially constitute a credible commitment to the particular strategy employed by the algorithm,<br><br>Speed: Fast detection and punishment of deviations from collusive prices by algorithms could stabilize collusion. Speed might also reduce the duration and cost of price wars, should they ever happen.<br><br>Behavioural consistency: The lack of human bias in algorithmic decisions is argued to facilitate coordination via increased predictability, but might also have the opposite effect as factors like guilt aversion might stabilize human cooperation.<br><br>Symmetry: The authors argue that algorithmic symmetry is insufficient for coordinated outcomes, especially when firms' market positions are not symmetric. |

---

[275] Gal, Michal S. "Algorithms as illegal agreements." *Berkeley Tech. LJ* 34 (2019): 67.